# The China Trade Shock and the ESG Performances of US firms

Hui Xu
Department of Accounting and Finance
Lancaster University
Email: h.xu10@lancaster.ac.uk

And

Yue Wu
Department of International Economics
Beijing Foreign Studies University
Email: yuewu@bfsu.edu.cn

How does import competition from China affect engagement on ESG initiatives by US corporates? On the one hand, reduced profitability due to import competition and lagging ESG performance of Chinese exporters can disincentivize US firms to put more resources to ESG initiatives. On the other hand, the shift from labor-intensive production to capital/technology-intensive production along with offshoring may improve the US company's ESG performance. Moreover, US companies have incentives to actively pursue more ESG engagement to differentiate from Chinese imports. Exploiting a trade policy in which US congress granted China the Permanent Normal Trade Relations and the resulting change in expected tariff rates on Chinese imports, we find that greater import competition from China leads to an increase in the US company's ESG performance. The improvement primarily stems from "doing more positives" and from more involvement on environmental initiatives. Indirect and direct evidence shows that the improvement is not driven by the change in production process or offshoring, but is consistent with product differentiation. Our results suggest that the trade shock from China has significant impact on the US company's ESG performance.

We thank Dan Bernhardt, Vikas Raman, Greg Pawlina, Sarah Kroechert, Gerald Ward, Mahmoud Gad, Chelsea Yao, George Wang, and participants at Lancaster University.

Corporates and investors increasingly play up environmental, social and governance (ESG) initiatives. A survey by U.S. Chamber of Commerce in 2021 finds that over half of the U.S. companies publish voluntary sustainability and ESG reports outside of their SEC filings. Investors are also concerned about ESG engagement and performance of corporates. The assets under management of global ESG ETFs are around $225 billion in 2020 and, according to Bloomberg Intelligence, are expected to grow at 35 percent per annum, reaching $1 trillion by 2025.[1] The investor's growing attention to ESG and corporate social responsibility (CSR) is also reflected in Google search volume index (see Figure 1).[2] In light of the growing spotlight on ESG, exploring the reasoning behind firms' ESG practices has gained relevance. In this paper, we focus on the relation between foreign import competition and ESG performance of domestic firms. Specifically, we study the effects of trade shocks from China on ESG performance of US domestic firms.

As China transited towards a market-oriented economy and reduced barriers to foreign trade during 1990s and 2000s, its share of world manufacturing exports grew from 2.3% in 1991 to 18.8% in 2013 (Autor, Dorn, and Hanson (2016)). Increasing import competition from China has profound impacts on the US economy, including survival of US manufacturing plants (Bernard, Jensen, and Schott (2006)), industry labor employment (Acemoglu et al. (2016)), labor incomes (Autor et al. (2014)), R&D expenditure and innovations of US firms (Autor et al. (2020)) and even political ideology (Autor, Dorn, and Hanson (2020)). However, it is not clear how import competition from China affect ESG engagement by local US firms. In fact, the theory and current evidence often leads to ambiguous and even contradictory conjectures.

On the one hand, facing greater competition from Chinese exporters who lagged behind on ESG performance, US firms may have less incentive to place resources into ESG initiatives. Furthermore, reduced profitability and tighter cash flow due to the competition can force firms to scale back on investment, R&D and ESG engagements. On the other hand, there are also reasons to believe that the ESG performance of US firms may improve as a response. For one, in light of the import competition from China, US firms increasingly shift to capital-and-technology intensive production segment and offshore manufacturing production operations overseas. The change in production process, therefore, can result in better ESG performance.

---

[1] According to the Global Sustainable Investment Association, global ESG assets surpass $35 trillion in 2020, up from $30.6 trillion in 2018 and $22.8 trillion in 2016.
[2] Da, Engelberg, and Gao (2011) show that Google search volume captures the investor's attention in a more timely fashion and more likely measures the attention of retail investors.

For another, given that Chinese exporters enjoy low cost of labor but lag behind on commitment to ESG initiatives, US firms may find it effective to engage more on ESG initiatives and differentiate from Chinese competitors. Therefore, the effect of import competition on corporate ESG engagement is a question that remains to be answered.

In this paper, we shed light on this question by exploiting the change in expected tariff rates on Chinese imports across industries as US congress granted Permanent Normal Trade Relations (PNTR) to China in 2001. US imports from nonmarket economies such as China are subject to non-Normal Trade Relations tariff (non-NTR) originally set under the Smooth-Hawley Tariff Act of 1930. Since 1980, imports from China to US were granted a waiver and enjoyed low NTR tariff rates. Such a waiver, nevertheless, was granted on an annual basis and had to be reviewed and re-approved by the US congress, causing an uncertainty of "the sword of Damocles". The PNTR permanently reduces the expected tariff rates and removes the uncertainty. Pierce and Schott (2016) shows that following PNTR, US producers experience increased import competition from China.

Following Pierce and Schott (2016), we exploit the variation in reduction of expected tariff rates across industries and adopt a difference-in-differences (DiD) approach. We find a strong positive relationship between reduction in expected tariff rates and ESG engagement by US firms. A standard deviation decrease in expected tariff rates raises a firm's ESG score by 0.48, approximately 21% of a standard deviation of the ESG score. The result suggests that firms in industries facing greater competition from China following PNTR substantially improve their ESG performance. In addition, our analysis shows that the improvement primarily stems from "doing more positives" and from more involvement on environmental initiatives. Our baseline analysis is robust to the inclusion of firm-level fixed effects, year fixed effects, and, more importantly, industry-year fixed effects.[3]

We proceed to study if the improved ESG performance is driven by changes in the production process. The literature in international trade, e.g., Mion and Zhu (2013), Autor, Dorn, and Hanson (2016) and Bloom, Draca, and Van Reenen (2016), shows that increased import competition from China has driven domestic producers to reduce employment, deepen capital intensity and offshore some manufacturing production. The change in production process can affect a firm's ESG performance. For instance, if firms hire less employees than before, then they may afford to offer higher salary and more benefits on a per capita basis, potentially leading to better employer-employee relations. To control for this, we further include variables related to the production process, both at the firm level and the industry level, in our specification, including R&D, number of staff, and capital intensity. The results show that these variables controlling for the production process account for little variation in firms' ESG performance, and have no explanatory power on the positive relation between reduction in expected tariff rates and ESG engagement by US firms, suggesting that the improvement in ESG performance is unlikely driven by changes in the production process.

---

[3] As Gillan, Koch, and Starks (2021) point out, ESG scores have a strong industry component. Industry-year fixed effects therefore can control for the unobservable industry factors that might be correlated with ESG. For instance, high pollution-producing and carbon-emitting industries may have more incentives over time to increase their ESG engagement due to pressure from regulators and investors.

Instead, we find both indirect and direct evidence supporting that US firms strategize ESG engagement to differentiate their products. Our first indirect evidence exploits the fact that firms in democratic-leaning ("blue") states on average have better ESG performance than those located in republican-leaning ("red") states (Di Giuli and Kostovetsky (2014)). Therefore, if ESG engagement is employed as a strategy upon greater import competition from China, firms in the "red" states, which lag behind on ESG performance than their peers in "blue" states, will display greater improvement, and this is indeed corroborated by our triple-difference test. Our second indirect evidence explores the relationship between the firm's market power and their ESG engagement. Using two different measures of market power (Hoberg and Phillips (2010); Hoberg and Phillips (2016)), we find that firms with less market power are more incentivized to improve their ESG performance upon trade shocks from China, consistent with the differentiation hypothesis.

If US firms indeed attempt to differentiate themselves from Chinese imports by actively engaging in ESG initiatives, firms producing standardized goods should have more incentives to do so. Our direct evidence therefore takes advantage of the product similarity measures from Hoberg and Phillips (2016) and from Rauch (1999), respectively and tests this insight. Indeed, the results from our empirical tests show that firms in standardized industries improve their ESG performance to a greater extent. Moreover, it also suggests that most of improvement on ESG performance documented in the baseline results come from companies operating in standardized industries. To our knowledge, our analysis is the first to present direct evidence of ESG engagement as a firm's differentiation strategy.

Our paper first contributes to a fast-growing literature on exploring the characteristics of firms and markets that could explain firms' ESG decisions.[4] Cai, Pan, and Statman (2016) and Liang and Renneboog (2017) provide evidence that country characteristics are important in explaining the firm's ESG activities. Di Giuli and Kostovetsky (2014) and Jha and Cox (2015) show that political leaning and social capital of the region in which a firm is headquartered also affect its ESG engagement. Studies also observe that a firm's ESG engagements are heavily influenced by the personal traits of its CEO and board of directors, including their genders, the genders of their children, and their marital status (Iliev and Roth (2021); Borghesi, Houston, and Naranjo (2014); Cronqvist and Yu (2017); Hegde and Mishra (2019)). In addition, Dimson, Karakaş, and Li (2015), Dyck et al. (2019), and Chen, Dong, and Lin (2020) among others, find that institutional investors can also exert significant influence on a firm's ESG engagement. Starks, Venkat, and Zhu (2017), however, argue that long-term institutional investors are in fact attracted to firms with higher ESG/CSR profiles instead of influencing the firm's choices directly. We extend this strand of literature by examining the effect of increased import competition on firms' ESG performance and further investigate the underlying mechanism. We provide novel evidence, both direct and indirect, showing the enhanced ESG engagement as a differentiation strategy for US producers upon greater competition from China.

Our paper is also related to an emerging literature that studies the impacts of rising import competition from China on US domestic markets. Much of the literature (David, Dorn, and Hanson (2013), Autor et al. (2014), Acemoglu et al. (2016) and Autor, Dorn, and Hanson (2019)) focus on the impacts of the trade shock on US labor markets. Autor et al. (2020) shows that US patent production declines in sectors facing greater import competition, while Autor, Dorn, and Hanson (2020) suggest that import competition from China may have contributed to the polarization of US politics. In this paper, we contribute to the literature by studying the impacts of trade shock from China on the ESG engagement of local US firms and analyzing the reason behind.

The paper proceeds as follows. Section I elaborates two competing hypotheses regarding the relationship between import competition from China and the ESG performance of local US firms. Section II introduces the background of US granting PNTR to China, and describes the data and our baseline identification strategy. Section III presents the main results.

---

[4] See Gillan, Koch, and Starks (2021) for a thorough review.

Section IV studies the mechanisms underlying the improvement in ESG performance, and showing that the improvement is not driven by the change in production process, but is consistent with ESG engagement being a differentiation strategy. Section V provides robustness checks, in which we use a different measure of trade shocks from China. Section VI concludes.

## I. US Firm's ESG Performance and Trade Shock from China

The trade shock and import competition from China has profound impacts on US firms, markets, community and even political ideology. However, it is not clear how the exposure to trade shock from China shapes ESG engagement and performance by US domestic firms.[5] In fact, competing theories suggest that foreign competition, e.g., trade shock from China, can increase or decrease the domestic firm's ESG engagement, leaving it to empirical studies to ascertain.

Import competition from China can decrease a US firm's engagement on ESG issues. The decrease can be ascribed to two effects: peer effect and cash flow effect.

Corporates pay close attention to their competitors, and an individual firm's decision-making is thus affected by its peers.[6] This also includes a firm's ESG engagement. Cao, Liang, and Zhan (2019) finds that an adoption of ESG proposal by a firm is followed by the adoption of similar ESG practices by peer firms. At the onset of China's transition towards a market-oriented economy and integration into world trade in 1990s and 2000s, Chinese firms enjoyed very low cost of labor, and did not prioritize ESG engagement, falling behind US firms during the same period. Anecdotal evidences suggest that the difference on ESG performance between Chinese firms and US firms may be significant.[7] Therefore, when competing with Chinese

---

[5] In this paper, we use ESG engagement and ESG performance interchangeably.
[6] Peer effects in corporate finance and governance are prevalent: Leary and Roberts (2014) argue that peer effect is a crucial determinant in firm's capital structure. Similarly, Grieser et al. (2021) find strategic complementarity in capital structure decisions. Kaustia and Rantala (2015) examine the effect of social learning and claim that firms are more likely to split their stock following their peers' actions. A number of studies show that peer effect plays an important role in chief executive officer (CEO) compensation, where corporations use their peer companies as benchmarks in determining the compensation packages (Faulkender and Yang (2010); Bizjak, Lemmon, and Nguyen (2011); Albuquerque, De Franco, and Verdi (2013)). Kelchtermans, Neicu, and Teirlinck (2020) and Peng, Lian, and Forson (2021) present evidence on imitation of firms in their R&D decisions and the usage of R&D tax exemptions. Foucault and Fresard (2014) and Dessaint et al. (2019) also find significant association between peer evaluation and corporate investment decisions.
[7] Hasanbeigi et al. (2014) compare the energy use and intensity between U.S. and China, and find that Chinese steel industry has much higher energy intensity than the U.S. in 2006, which has a direct impact on energy consumption and related carbon dioxide ($CO_2$) emissions. The Carbon Disclosure Project (CDP5) in 2007 highlights the overall lack of data from Chinese corporates. In addition to environmental issues, there are also

firms, US firms of which the ESG performance already stood out may choose to spend more on other value-creating activities, e.g., R&D, instead of putting more resources into ESG engagement. Furthermore, ESG rating likely reinforces the peer effect on ESG engagement. While some rating agencies score firms relative to the entire firm universe, others use industry benchmarking, scoring a firm relative to other firms in the similar line of business (Gillan, Koch, and Starks (2021)).[8]

The peer effect can occur independent of a firm's cash flow level. However, foreign competition can also squeeze a firm's profitability and tighten its cash flow (Esposito and Esposito (1971), Pugel (1980)), further reducing a firm's ESG engagement on top of the peer effect. The tight cash flow, in principle, is expected to downsize both a firm's value-creating projects, e.g., R&D and investment (Autor et al. (2020)), and other activities, e.g., charity and donations (Gregory, Tharyan, and Whittaker (2014)). ESG engagement is no exception: Hong, Kubik, and Scheinkman (2012) show that financial constraint adversely affects a firm's engagement in ESG issues.

While peer effect and cash flow effect may lead to lower ESG performance in the presence of trade shock from China, other factors can incentivize a firm to engage more on ESG issues, either passively or actively.

The first factor is the change in production process caused by the trade shock. Pierce and Schott (2016) show that US producers reduce employment and deepen capital intensity when facing greater competition from China following the grant of PNTR to China. Such change in the production process may inadvertently improve a firm's ESG performance. For instance, if a firm hire less employees than before, then they may afford to offer higher salary and more benefits on a per capita basis, leading to better labor relations. In fact, despite the significant unemployment related to trade shock from China, Figure 2 shows that the job satisfaction in the US stayed steady or even ticked up in the 2000s after the PNTR was passed, suggesting potential differential impacts of the change in production process on people staying

---

concerns about the labor right protection in China. China Labor Watch, an independent not-for-profit organization, conducts assessments of factories in China on their labor conditions. Their reports over the years have shown that many Chinese factories experience problems such as employment of underaged workers, high work hours, gender discrimination and so on. In one of their 2007 reports that receives media attention, they investigate toy suppliers in China and claim that workers are suffering brutal conditions and illegal practices. (China Labor Watch, 2007 Aug 21, retrieved from https://chinalaborwatch.org/investigations-on-toy-suppliers-in-china-workers-are-still-suffering/)

[8] For example, MSCI ESG Rating assess "company Risk Exposure and Risk Management relative to industry peers" and creates "an overall ESG rating (AAA – CCC) relative to industry peers".

in the job and people out of the job. Most of the ESG rating agencies also leave out firms' decision on the termination of employment in their rating methodology, and only focus on the welfare policies of employees that are currently on the job. Relatedly, US firms have increasingly moved part of their operation offshore to China. Recent evidence shows that corporates in the developed economies where environmental policies are rigorous offshore production with high pollution and high $CO_2$ emission to regions where environmental protection is weak, engaging regulatory arbitrage. [9] Therefore, one cannot rule out *a priori* that the improvement in US firms ESG performance stems from offshoring part of their production process to China.

Secondly, a firm may improve its ESG performance to differentiate itself from the Chinese competitors. Management and marketing literature have argued that firms use better ESG performance to differentiate themselves (Navarro (1988); Bagnoli and Watts (2003); Siegel and Vitaliano (2007); Hull and Rothenberg (2008); Flammer (2015)) and to foster customer loyalty. Creyer and Ross (1997) show that consumers care about firms' ESG performance and take it into account when making purchasing decisions. Auger et al. (2003) and De Pelsmacker, Driesen, and Rayp (2005) indicate that consumers are willing to pay for better ESG feature of a product. Therefore, US domestic firms may have great incentive to improve ESG performance facing greater competition from Chinese exporters who enjoy lower cost of labor but lag behind on ESG engagement. The incentive to improve, however, might differ depending on the firm's headquarter location, market power and the characteristics of the industry in which the firm operates, which we will detail in Section IV.

In summary, theoretic analysis offers opposite propositions on how import competition from China affects US firms' engagement on ESG issues. Even within the same proposition, theories have suggested different possible channels. Therefore, we turn to empirical analysis to shed light on the impact of the trade shock and the potential mechanisms.

---

[9] Ben-David et al. (2021) find that firms allocate their pollutions internationally driven by environmental policies in the home country. Moran, Hasanbeigi, and Springer (2018) investigate global carbon trade and estimate that 25% of the global carbon emissions can be account for by production offshored abroad. Hasanbeigi, Morrow, and Shehabi (2021) take a closer look at U.S. manufacturing and trade, and detect high level of global carbon footprint embodied in traded goods in the U.S. Various news articles make similar claims and bring attention to pollution outsourcing to developing countries in the global supply chain. See, for example, the Guardian (https://www.theguardian.com/environment/2014/jan/19/co2-emissions-outsourced-rich-nations-rising-economies) ; the New York Times (https://www.nytimes.com/2018/09/04/climate/outsourcing-carbon-emissions.html).

# II. Data

**Uncertainty and Expected Tariff Rates on Imports from China:**

Our primary identification strategy exploits a US trade policy granting Permanent Normal Trade Relations (PNTR) to China in 2001, following Pierce and Schott (2016).[10] Imports from China to the US had been subject to the relatively low NTR tariff rates reserved for WTO members since 1980. These low rates, however, required annual reviews and approval by Congress. Had Congress not renewed China's NTR status, the tariff rates on imports from China would have become non-NTR rates originally set by the Smoot-Hawley Tariff Act of 1930 and significantly hiked. To see the difference between NTR and non-NTR rates, the average tariff rates on Chinese imports is 3.4% in 1999, and this figure would have jumped by 10 times to 37% without NTR status.

The renewal process by the US congress was not a bureaucratic formality. In fact, every year between 1990 and 2001, the US House of Representatives brought and voted on the bill attempting to revoke the China's temporary NTR status. The uncertainty and potential increase in expected tariff rates were finally removed by PNTR.

To gauge the import competition from China across industries, we use the *NTR gap* calculated by Pierce and Schott (2016). Specifically, NTR gap for industry $j$ is the difference between the non-NTR rates that would have applied to imports in industry $j$ from China had the annual reauthorization failed, and the NTR tariff rates set by PNTR,

$$NTR\ Gap_j = Non\ NTR\ Rate_j - NTR\ Rate_j$$

Higher *NTR gap* in industry $j$ indicates more intense competition from Chinese imports following PNTR. Pierce and Schott (2016) find that US industries with higher gaps between non-NTR and NTR tariffs experience acceleration in Chinese imports. They also show that 79% of the variation in NTR gap comes from the variation in non-NTR rates that were set in the 1930s. This feature renders NTR gap plausibly exogenous to ESG engagement by US firms after 2001. Moreover, the first-order consideration concerning the legislation in the tariff and trade agreement is comparative advantage in cost of production of both domestic and foreign firms and local labor employment (Gros (1987); Elhanan and Krugman (1989)), probably not the ESG performance of the firms.

---

[10] The legislation was passed by the House of Representatives on May 24, 2000 and by the Senate on September 19, 2000. The President signed on Oct 10, 2000. It officially became effective when China joined the WTO in December, 2001.

The *NTR gaps* from Pierce and Schott (2016) are set at eight-digit Harmonized System (HS) level. To concord these data to four-digit SIC level, we use the crosswalk provided by David, Dorn, and Hanson (2013) which slightly aggregate the four-digit SIC industries so that each of the resulting manufacturing industries matches to at least one HS code. We then take the simple average of NTR gaps at four-digit SIC level. Similar to Pierce and Schott (2016), we use the *NTR Gaps* from 1999, but will show in Section V the robustness of our results to *NTR Gaps* from other years.

**ESG Engagement:**

Our data on ESG engagement by US firms come from Kinder, Lydenberg and Domini (KLD) Research & Analytics, Inc. KLD started to collect the scores of ESG engagement in 1991 for 488 firms, and the coverage grew over the years to include 2894 firms in 2009. After 2009, the calculations of ESG scores changed (Hong et al. (2019)). Considering the 2008 global financial crisis and its potential impacts on firms' ESG commitment, our analysis uses the KLD information from 1991 to 2007.[11] The time span of our analysis also coincides with those studying the impacts of rising Chinese import on US labor markets and firm innovations, e.g., Acemoglu et al. (2016).

The KLD ratings are built on a point-by-point assessment of companies along a number of dimensions. Firms are graded on roughly 60 indicators. Each indicator represents a strength or a concern in one of six major areas: environment, community, diversity, employee relations, product, and corporate governance. A firm gets a score of 1 for a strength (concern) indicator if it performs well (poorly) in a particular criterion, and zero otherwise. For instance, Table 1 shows two indicators of strengths and two indicators of concerns associated with the area of diversity. If a firm has strong gender diversity on board of directors and among executive management team, the firm would score 2 on the strength of diversity, and at most score 1 on the concerns of diversity.

We measure a firm *i*'s overall ESG performance in year *t* as the difference between the sum of total strengths and the sum of total concerns during year *t*. Similarly, we take the difference between the number of strengths and concerns of each constituent environmental, social, and governance factor as measures for a firm's performance on E, S, and G initiatives, respectively. KLD also includes a set of indicators regarding human rights, an assessed area that contributes to a firm's performance on "S" initiative. However, as noted in Hong, Kubik, and Scheinkman (2012), the area went through a major overhaul in 2002 and is therefore not consistent throughout our sample period. In addition, KLD tracks controversial business involvement related to alcohol, firearms, gambling, military, nuclear power, and tobacco. These indicators, nevertheless, are very specific to a small number of firms, do not apply to most of the firms in our sample, and, hence, are also excluded from our ESG measures. More detailed categories and indicators included in our ESG measures are tabulated in Table A.1 in the appendix.

---

[11] As far as we are aware, KLD is the only data that started to record a firm's ESG/CSR activities in 1990s. Other data, e.g., Thomas Eikon (2001), Sustainlytics (2008), MSCI ESG (2007) and RepRisk (2008), started in 2000s.

**Firm-level controls**

Following Di Giuli and Kostovetsky (2014), we collect a number of time-varying firm-level controls from Compustat and CRSP, and include them in all of our empirical analysis. For each firm-year, we control for the firm's size measured by (log) total assets, return to assets (ROA), book-to-market ratio, cash, dividends, and total debt outstanding (leverage ratio). All control variables are lagged by one year. In addition to the time-varying firm-level controls, we also include firm fixed effects to control for time-invariant heterogeneity across firms that might affect ESG performance, and include year fixed effects to control for time-varying common shocks to ESG engagement of all firms in the sample. Detailed definitions and construction of the control variables are in Table A.2 in the appendix.

Gillan, Koch, and Starks (2021) highlight a strong industry component of ESG scores. While the firm and year fixed effects mitigate the heterogeneity across industries to some extent, they still have shortcomings. In particular, year fixed effects do not absorb industry factors likely correlated with ESG, and firm fixed effects presume no temporal variation in industry-specific unobservable characteristics that may bias the results. If, for instance, "brown" industries have more incentives over time to increase their ESG engagement due to regulators' pressure or investors' preferences, then failing to account for heterogeneity across industries would skew the results. Therefore, we also include industry-year fixed effects to account for different trends in ESG engagement across industries.

Table 2 displays the summary statistics of ESG scores, individual constituent scores, and other controls for the sample firms from 1991 to 2007. The mean ESG score is slightly negative at -0.10, indicating that the number of concerns exceeds the number of strengths. But it displays a large variation with a standard deviation of 2.32. Among individual constituent scores, US firms excel at social issues with the mean S score being 0.3. Nevertheless, S score also has a greater standard deviation relative to E and G scores, suggesting performances on social issues across different years and firms contribute to a great proportion of the variations in overall ESG scores.

## III. Results

We study the impacts of import competition from China on US firms' ESG engagement by adopting a difference-in-difference (DiD) approach. The first difference in our DiD strategy is between NTR gaps across 4-digit SIC industries. The second difference is between the period before the passage of PNTR by the US congress in 2001 and the period after. We estimate the following equation:

$$ESG_{i,j,t} = \alpha Post\_PNTR_t \times NTR\ Gap_j + \beta X_{i,t} + \eta_t + \delta_j \times \eta_t + v_i + \epsilon_{i,t} \qquad (1)$$

where the dependent variable $ESG_{i,j,t}$ is the ESG score of firm $i$ in industry $j$ during year $t$. $Post\_PNTR$ is an indicator variable equal to 1 if year $t$ is after 2001 and 0 otherwise. $NTR\ Gap_j$ is the difference between non-NTR and NTR tariff rates observed in 1999 for industry $j$. $X_{i,t}$ is the vector of firm-level covariates; $\eta_t$ represents year fixed effects; $\delta_j$ is industry fixed effects, and thus $\delta_j \times \eta_t$ controls for different trends in ESG performance across industries; $v_i$ indicates firm fixed effects. The coefficient $\alpha$ of the interaction term thus captures the impacts of greater import competition, following China being granted with PNTR, on the ESG performances of US firms.

Table 3 shows the results of our baseline regression. Through all different specifications, we find evidence that greater import competition from China *increases* the ESG engagement of US firms. Column 1, 2 and 3 include full covariates, firm fixed effects and/or year fixed effects, and show significant coefficients between 1.3 and 3.1. Based on the result in Column 3, a standard deviation increase in *NTR Gap* (0.15) approximately raises a firm's ESG score by 0.47. This is about 21% of a standard deviation (2.32) of ESG scores, and 37% of a standard deviation (1.25) of the yearly change in ESG scores.

ESG scores have strong industry-specific components. Industries such as mining and chemicals have lower scores than industries such as management consulting, banking and insurance. In addition, this industry variation may also change over time and correlate with NTR gaps. Although our primary focus on manufacturing industries along with the inclusion of firm and year fixed effects alleviates the concern to some extent, to further control for the heterogeneity in ESG engagement across industries over time, we include industry (three-digit SIC level)-year fixed effects in Column (4). Relative to Column (3), the coefficient of interest in Column (4) falls by 38%, highlighting the importance of accounting for heterogeneity across industries. Nevertheless, the coefficient is still statistically significant at 1% level, suggesting a standard deviation rise in *NTR gap* increases a firm' ESG score by 0.29, roughly 12% of a standard deviation of ESG scores.[12]

There are many ways for firms to improve their ESG scores. They can increase the scores on strengths by doing more positives. Alternatively, they can reduce the scores on concerns by amending the negatives. Moreover, firms can also select among environmental, social or governance-related initiatives to engage in, and enhance performance on one or multiple constituents of ESG scores, all resulting in better overall ESG performance.

Table 4 studies the probable source of the ESG improvement documented in Table 3. Column (1) and (2) study the impacts of passage of PNRT on scores of strengths and concerns using the most saturated specification of (1). The results show that the scores on strengths significantly increase after the passage of PNRT while the scores on concerns barely change, suggesting that the ESG improvement in Table 3 is primarily from "doing more positives" by the firms. Column (3) to (5) study scores of each E, S and G component. The coefficient of the interaction term for environmental issues is 1.02, stand close to the coefficient for overall ESG performance in Table 3, and is significant at 1% level. In contrast, the corresponding coefficients for social and governance-related scores are not significant at the conventional levels. This suggests that the better ESG performance in Table 3 is mainly driven by firms engaging more in environment-related activities, e.g., reducing toxic emissions and waste.

**Identifying Assumptions**

---

[12] We also estimate a specification only including firm, year and industry-year fixed effects, but without firm-level controls. The estimated coefficient is 2.236, and significant at 1% level. The fact that the coefficient is much smaller than Column (3) and is close to Column (4) with full firm controls suggests that the variation in ESG engagement mostly clusters at industry level.

The underlying assumption of our DiD strategy is that industries exposed to different *NTR gaps* after 2001 should have similar trends in ESG performance beforehand. In order to test the parallel-trend assumption prior to the passage of PNTR and ensure the validity of estimation strategy, we repeat the specification in (1) but replace the *Post_PNTR* indicator with a series of year indicator variables:

$$ESG_{i,j,t} = \sum_{n=1991}^{2007} \alpha_n \times Year_{n,t} \times NTR\,Gap_j + \beta X_{i,t} + \eta_t + \delta_j \times \eta_t + v_i + \epsilon_{i,t} \qquad (2)$$

where $Year_{n,t}$ is a dummy variable equal to one if $n=t$, and zero otherwise. To satisfy the underlying assumption, the coefficients $\alpha_n$ for $n$ before 2001 should not be significant. In addition, the series of coefficients $\{\alpha_n\}$ also capture the dynamic effects of granting PNTR to China on US firms' ESG engagement. As it takes time to deploy resources to improve ESG performance, we hypothesize that the change on ESG engagement is gradual rather than instant, and becomes more significant over time.

Figure 3 plots the coefficients and their 95% confidence intervals for the specifications without firm-level controls (solid line) and with firm-level controls (dashed line). For both specifications, the coefficients prior to 2001 are not statistically significant, supporting the parallel trend assumption of our DiD analysis. It also shows that the improvement of ESG performance of US firms is a gradual process and becomes more evident over time, confirming our initial hypothesis.

## IV. Mechanisms

Having established that trade shocks from China increase US firms' ESG engagement, we try to shed light on the possible channels in this section. As discussed in Section I, two possible channels are at play: ESG performance can improve as a consequence of adjusting production process or product mix in face of greater competition, or offshoring some production operation to China; alternatively, US firms may seek to differentiate themselves from Chinese imports by actively engaging in more ESG initiatives. We find no evidence to support the first mechanism. However, we find both indirect and direct evidence supporting the hypothesis of product differentiation.

## A. ESG Improvement: A Consequence of the Change in Production Process?

The literatures studying the impacts of Chinese import competition have found change in the production process or product mix among US firms (Autor et al. (2014); Pierce and Schott (2016); Bloom, Draca, and Van Reenen (2016); Autor et al. (2020)). When faced with increased import competition from China, US firms typically downsize employment, reduce R&D expenditure, and deepens capital intensity. Do such changes in production process result in better ESG performance? In other words, firms may not actively pursue superior ESG performance, but are only passively rated higher.

To study whether the increase in ESG performance of US firms is driven by change in production process, we further include four variables related to production process in our baseline regression (1). The first variable is staff expense scaled by sales, and it captures the change of labor share in the production process. The second variable is capital intensity measured as the capital expenditure scaled by the total number of employees. The third and fourth variables are expenditure on R&D and advertising scaled by sales, and they measure firms' investment in brand name and intellectual capital. In addition, we also include the interaction terms between the four variables and *post_PNTR* indicator to account for potential change in the relationship between ESG scores and these characteristics in the post PNTR period. If the ESG improvement we find in Table 3 is indeed a result of change in production process, these variables should capture a significant proportion of variations in ESG scores, both cross-sectionally and temporally. Consequently, the coefficient of DiD term should also decrease significantly and even become statistically insignificant.

Table 5 presents the explanatory power of these additional variables. When compared with Table 3, it shows that the variables related to production process can only account for a minuscule proportion of the increase in ESG engagement, if any. None of the additional controls and their interactions with *post_PNTR* show statistical significance at the conventional levels. Even with the inclusion of all four variables in Column (5), the coefficient of the DiD term is still 1.83 and significant at 1% level. This suggests that the change associated with production process is unlikely to account for higher ESG scores of US firms after the passage of PNTR.

It is well known that a significant proportion of firms in Compustat have missing values in staff expense, R&D expenditure and advertising expenditure (Donangelo et al. (2019)). Our sample is no exception. Only 16% of the firm-year observations have firm-level R&D, staff and advertising expense in the Compustat. Following a large literature (e.g., Fee, Hadlock, and Pierce (2009) and Masulis, Wang, and Xie (2009)), we set the missing values to zero. This might affect our estimations in Table 5. To mitigate this issue, we compute the average R&D, staff, advertising expenditure and capital intensity at the 4-digit SIC industry level each year and re-run the regression.

For the industry-level analysis, we are also able to control for production offshoring by US companies. Decreased trade barrier means that firms can have access to lower cost intermediate inputs from China, and offshore part of their production process to China. If the firm primarily keep more technologically-advanced and eco-friendly production at home, this could lead to increase in ESG performance. We study this mechanism and control for the firm's offshoring to China following Bloom, Draca, and Van Reenen (2016). For each industry $j$, we compute an offshoring measure adapted from Feenstra and Hanson (1999):

$$Offshoring_{j,t} = \sum_{k} w_{j,k} IP_{k,t} \qquad (3)$$

where the input-output weight $w_{j,k}$ measures the weight of inputs in industry $k$ needed to produce one unit of final good in industry $j$. $IP_{k,t}$ is import penetration from China in US industry $k$ and defined as:

$$IP_{k,t} = \frac{M_{k,t}^{UC}}{Y_{k,91} + M_{k,91} - E_{k,91}} \qquad (4)$$

$M_{k,t}^{UC}$ is the imports from China in industry $k$ during year $t$; $Y_{k,91} + M_{k,91} - E_{k,91}$ is the initial absorption level at the start of the period in 1991, with $Y_{k,91}$, $M_{k,91}$, and $E_{k,91}$ representing shipments, aggregate imports, and aggregate exports in industry $k$, respectively. We collect the input-output weights from US Bureau of Economic Analysis (BEA), and the import penetration variable from Acemoglu et al. (2016).[13]

Table 6 shows the industry-level result. Notably, Column (1) and (2) show that staff expense and capital intensity can account for some variation in ESG scores. After 2001, increases in staff expense and capital intensity is positively associated with better ESG performance. Nevertheless, controlling for variables related to production process at the industry level only increase the magnitude of the coefficient of the DiD term, if anything. Column (6) include all five variables together, and the coefficient of interest is 3.72 and significant at 5% level, suggesting one standard deviation increase in NTR gap raises the ESG scores by 0.56 post PNTR. This translates to 24% of one standard deviation of ESG scores. The results in Table 6, therefore, further suggest that better ESG performance of US firms is unlikely driven by the change in production process or offshoring.

---

[13] Note that the input-output weights at the most detailed industry level are not available every year. We use the weight data from 1997. The input-out matrix is defined in BEA industry classification. We first match it to NAICS using concordance table provided by BEA, and then further match it to SIC with the crosswalk table from David, Dorn, and Hanson (2013). For each BEA industry matched to multiple SICs, we divide the weight by the number of SICs it is matched to. For multiple BEA industries matched to one SIC, we sum over the weights of the BEA industries for the SIC.

### B. ESG Improvement: An Effort to Differentiate? Indirect & Direct Evidence

In this subsection, we present indirect and direct evidence suggesting US firms actively pursue better ESG performance and differentiate themselves from Chinese exporters.

**Headquarter Locations, Politics and ESG Engagement**

Di Giuli and Kostovetsky (2014) find that firms headquartered in Democratic party-leaning states are more likely to spend resources on ESG and on average have better ESG performance than those headquartered in Republican party-leaning states. If firms attempt to contrast themselves from the imports from China by ESG engagement, firms located in Republican-leaning state will likely display greater improvement in ESG engagement in light of the greater import competition. We therefore test this insight as our first indirect evidence showing the increase in ESG engagement is an effort of product differentiation.

We collect headquarter states of the sample firms from Compustat and their historical 10K filings.[14] To measure party affiliation and party strength across states, we follow the literature and consider four measures including the proportion of the Congress delegation that is Republican, an indicator variable ("*Red_congress*") equal to 1 if the percentage of Republican delegates is greater than Democratic, the average percentage of electorates in the state who voted for the Republican candidate, and an indicator variable ("*Red_president*") equal to 1 if the average percentage of electorates voting for the Republican candidate is greater than the average percentage of electorates voting for the Democratic candidate. Given our focus on periods between 2001 and 2007, our measures are constructed using presidential and midterm congressional elections between 2000 and 2006. To study if firms headquartered in republican-leaning states will make a greater effort to increase ESG performance to differentiate themselves, we consider a triple-difference specification:

$$ESG_{i,j,t} = \alpha_1 Post\_PNTR_t \times NTR\ Gap_j \times Republican_i + \alpha_2 \times Post\_PNTR_t \times NTR\ Gap_j \\ + \alpha_3 \times Post\_PNTR_t \times Republican_i + \beta X_{i,t} + \eta_t + \delta_j \times \eta_t + v_i + \epsilon_{i,t} \quad (5)$$

where $Republican_i$ denotes the Republican party affiliation and strength of firm $i$'s headquarter state using our measures. If firms located in Republican-leaning state display a greater increase in ESG engagement when faced with import competition from China, the coefficient $\alpha_1$ should be significantly positive.

---

[14] The headquarter location is a header variable in Compustat. In other words, Compustat always shows a firm's most recent headquarter location. We use a firm's historical 10K filings to track any change in headquarter locations over time.

Table 7 presents the result. Column 1 uses the proportion of Republican delegates in the Congress to measure the extent to which the headquarter state is Republican-leaning. Although the coefficient is not precisely estimated (p-value=0.14), it shows a positive sign as anticipated. When the leaning toward Republican is measured with the dummy variable *Red_congress* in Column (2), however, we see a positive coefficient significant at 5% level, suggesting that a firm in Republican states increase ESG engagement more than their peers in Democratic states. When we employ the alternative measures of the leaning to Republican with presidential elections in Column (3) and (4), the results become stronger. Both coefficients of the triple-difference are significant at 1% level, and the magnitude is also greater. This confirms that our finding is robust to different measures of party leaning, and that firms in Republican states improve their ESG performance more upon the passage of PNTR relative to their peers in Democratic states.

**Market Power and ESG Engagement**

Firms with great market power are less vulnerable to competition. In a competitive economy like the US, market power is likely to arise due to technical barrier, unique asset, and regulatory reason. Therefore, although Chinese producers enjoy low cost of labor, they can hardly pose threat to a local firm with high market power. Studies (e.g., Li, Lo, and Thakor (2021)) show that firms with market power are less motivated to innovate. In the same vein, firms with great market power should have less incentives to resort to ESG engagement to differentiate.

We gauge market power with two measures of Herfindahl–Hirschman index (HHI). The first measure is collected from Hoberg and Phillips (2010). Using Compustat data and actual industry HHI from the Commerce Department, Hoberg and Phillips (2010) estimate HHI for all 3-digit SIC industries. Our second measure is derived from Hoberg and Phillips (2016). Drew on text-based analysis of product descriptions from 10-K filings, Hoberg and Phillips (2016) identify a group of competitors for each firm in each year, and compute HHI for the firm based on the group. The grouping methodology is not required to be transitive and, as Hoberg and Phillips (2016) argue, "further benefit from information about the degree to which specific firms are similar to their competitors." We then take the average of firm-level HHI across 3-digit SIC industries. [15]

We take each industry HHI measure in 1999, and label industries of which the HHI are above the median with an indicator variable $High\_HHI$. We then consider a triple-difference specification similar to equation (3), including the triple interaction between $Post\_PNTR$, $NTR\ Gap$, and $High\_HHI$, the simple interaction, non-interacted terms and the same controls.

$$ESG_{i,j,t} = \alpha_1 Post\_PNTR_t \times NTR\ Gap_j \times High\_HHI_j + \alpha_2 \times Post\_PNTR_t \times NTR\ Gap_j$$
$$+ \alpha_3 \times Post\_PNTR_t \times \times High\_HHI_j + \beta X_{i,t} + \eta_t + \delta_j \times \eta_t + v_i + \epsilon_{i,t} \quad (6)$$

Table 8 reports the results. Column (1) and (2) use the measure of Hoberg and Phillips (2010). When including industry-year fixed effects, Column (2) shows a negative coefficient significant at 1% level, indicating that firms with greater market power are less likely to increase their ESG engagement in light of the import competition from China. Column (3) and (4) employ the measure from Hoberg and Phillips (2016), and display similar results. Depending on the specification, the p-value is between 0.04 (when industry-year fixed effects are not included) and 0.17 (when industry-year fixed effects are included). Moreover, throughout all measures and specifications, the results yield negative coefficients consistently, as conjectured.

---

[15] Both measures of HHI are directly downloadable from Hoberg-Phillips Data Library: https://hobergphillips.tuck.dartmouth.edu/

**Product Differentiation and ESG Engagement**

The evidence in Table 7 and Table 8 is consistent with the conjecture that firms in Republican-leaning states and with less market power make greater efforts to improve ESG performance to contrast themselves from the imports of China. Nonetheless, such evidence is circumstantial and merely suggestive. To provide direct evidence, we employ two measures of product differentiation.

Our first measure is derived from Hoberg and Phillips (2016). Similar to the HHI measure, Hoberg and Phillips (2016) also compute a similarity measure within a group of close peers by textual analysis using the product descriptions in 10-K filings. We take the average of firm-level product similarity measures in 1999 across 3-digit SIC industries, and designate industries with similarity measures above the median as standardized.

Our second measure follows Rauch (1999) and Giannetti, Burkart, and Ellingsen (2011) to define standardized and differentiated products. Commodities like unwrought lead which are traded on organized exchanges, and commodities like polymerization and copolymerization products that have reference prices in trade publications, can be regarded as standardized goods, as traders can solely base their profit estimates and import decisions on the reference prices without knowing the names of the manufacturers. On the contrary, goods like shoes, do not have reference prices, and local prices need to vary, for example, according to the varieties of local shoes and preferences of local consumers, and they therefore fall under the definition of differentiated goods. We use data on standardized and differentiated products from Rauch (1999) which classify the products at the 4-digit Standard International Trade Classification (SITC) level, and match the 4-digit SITC to 4-digit SIC.[16]

If US firms indeed attempt to differentiate from Chinese imports by actively engaging in ESG initiatives, firms producing standardized commodities should have more incentives to do so. Therefore, we consider the following regression,

$$ESG_{i,j,t} = \alpha_1 Post\_PNTR_t \times NTR\ Gap_j \times Standardized_j + \alpha_2 Post\_PNTR_t \times NTR\ Gap_j \\ + \alpha_3 Post\_PNTR_t \times Standardized_j + \beta X_{i,t} + \eta_t + \delta_j \times \eta_t + v_i + \epsilon_{i,t} \quad (7)$$

---

[16] The data is available from James Rauch's website: https://econweb.ucsd.edu/~jrauch/rauch_classification.html. We first match 4-digit SITC to 6-digit HS using the concordance table from the World Bank, and then match to 4-digit SIC code using the concordance table from David, Dorn, and Hanson (2013). One SITC code may correspond to multiple SICs, and vice versa. We then use the detailed US imports at 6-digit HS level from Schott (2008) and aggregate the classification to unique 4-digit SIC.

where $Standardized_j$ is a dummy variable that equals 1 if a firm is operating a standardized industry $j$. $\alpha_1$ captures the relative ESG engagement between standardized and differentiated industries upon greater trade shock from China.

Table 9 shows that is indeed the case. Throughout both measures and all specifications, the results yield positive estimates for $\alpha_1$. Column (1) and (2) use the similarity measure from Hoberg and Phillips (2016). Depending on the specification, the p-value is between 0.06 (when industry-year fixed effects are not included) and 0.11 (when industry-year fixed effects are included). Column (3) and (4) instead use the standardized industry classification from Rauch (1999).[17] When including industry-year fixed effects, Column (4) shows a positive coefficient significant at 1% level, suggesting that firms in standardized industries increase their ESG engagement to a greater extent than those in differentiated industries, as the theory suggested.

Moreover, comparing Table 9 to Table 8 and Table 3 also reveals some insights. Both Table 8 and Table 9 present results from triple-difference regressions. However, in Table 8, all the coefficients of the simple difference term, $Post\ PNTR \times NTR\ Gap$, are significant at the conventional levels, and their magnitudes are comparable to Table 3, suggesting that a substantial proportion of the variation in ESG engagement following PNTR is unaccounted for. In contrast, in Table 9, all the coefficients of the simple difference term are substantially smaller than Table 3, and except Column (3), the coefficients are either not or borderline significant (p-value is 0.10 in Column (2)) at the conventional levels. Hence, this further suggests that most of the increase in ESG engagement after 2001 comes from firms in standardized industries, supporting the notion that firms dedicate resources to ESG improvement in order to differentiate themselves.

In summary, we look at both indirect evidence (headquarter locations, market power and ESG engagement) and direct evidence (product differentiability and ESG engagement), all of which are consistent with the differentiation hypothesis of the ESG engagement.

---

[17] Rauch (1999) includes two ways to classify the products at the 4-digit SITC level: "conservative" and "liberal". The conservative classification minimizes the number of commodities being classified as standardized, while the liberal one maximizes the number. Table 9 reports the conservative classification. Table A.3 in the appendix shows that the result is robust if we adopt the liberal classification.

# V. Robustness Check

## V.1 Instrument for NTR Rates

Our baseline results in Table 3 use the *NTR Gap* in 1999. To assess the exogeneity of *NTR gap,* we also follow Pierce and Schott (2016) and instrument the constructed *NTR Gap* with two separate instruments: non-NTR tariff rates set by Smoot–Hawley Tariff Act, and the *NTR gap* observed in **1990**. Both of the instruments are distantly ahead of PNTR and the subsequent increase in US firms' ESG engagement, and thus, are plausibly exogenous.

Table 10 presents the second stage of IV regressions. As a reference, Column (1) reports the baseline results from Table 3. Both IV regressions yield similar results in terms of coefficient magnitude and statistical significance, indicating that moving *NTR Gap* from 25th percentile to 75$^{th}$ percentile increases a firm's ESG score by 0.311 (based on Column (2) result), approximately 13.4% of one standard deviation of the ESG score. Hence, the IV regressions further confirm that US firms improve their ESG performance following the Congress granting PNTR to China.

## V.2 Alternative Identification

Our main identification strategy follows Pierce and Schott (2016) by exploiting the change in trade policy related to China's attainment of PNTR in 2001, and by exploiting the variations across industries in expected tariff increase on imports from China. Although our robustness checks, e.g., the pre-trend analysis and IV estimations, strongly support the exogeneity of the DiD strategy, to further substantiate our findings, we follow Autor et al. (2014) and present an alternative identification strategy. Specifically, we use Chinese import penetration in the United States as the measure for trade exposure to Chinese goods across different U.S. industries. We then show that the rise in industry-level Chinese import penetration from year 1991 to 2007 leads to increase in U.S. firms' ESG performance.

As in Autor et al. (2014), the change in import penetration is defined as

$$\Delta IP_{j,t} = \frac{\Delta M_{j,t}^{UC}}{Y_{j,91} + M_{j,91} - E_{j,91}} \tag{8}$$

at each four-digit SIC industry level $j$, where $\Delta$ denotes the eight-year long difference operator over time period $t$. $\Delta M_{j,t}^{UC}$ is the change in imports from China, and $Y_{j,91} + M_{j,91} - E_{j,91}$ is the initial absorption level at the start of the period in 1991 similarly defined in equation (4). We consider the following regression model,

$$\Delta ESG_{i,j,t} = \alpha_1 \Delta IP_{j,t}^{CH} + \Delta X_{i,j,t} + \Delta \epsilon_{i,j,t} \tag{9}$$

where $\Delta ESG_{i,j,t}$ is the ESG score change of firm $i$ operating in industry $j$ over time period $t$; $\Delta IP_{j,t}^{CH}$ is the change in Chinese import penetration; $\Delta X_{i,j,t}$ are a set of change in firm level control variables. In essence, regression equation (9) is a first-difference estimator of a fixed-effect model with panel data. Unlike our primary identification strategy, the trade shock from China is now measured with import penetration.

We collect the industry-level import penetration data from Acemoglu et al. (2016), and match it to our firm-level ESG data. Our final sample consists of two stacked sub-periods, 1991 to 1999 and 1999 to 2007. Note that given KLD's expanding coverage of firms over time, a firm appearing in the sample in the later years might not show up in the sample in the early years. Therefore, a firm appears in the first period if it is covered by KLD in both 1991 and 1999; similarly, it appears in the second period if it is covered by KLD in both 1999 and 2007.[18] For this reason, the final sample size is substantially smaller than the one in the primary identification strategy.

One caveat is that the growth in Chinese import penetration may be related to unobserved shocks to US domestic productivity or demand, i.e., consumer preferences, that is also correlated with firms' ESG performance, resulting in potential biased estimates. To correct for the bias, we follow Autor et al. (2014) and instrument for the rising import penetration of China in the United States using the change in industry-level import penetration in other high-income countries,

$$\Delta IP_{j,t} = \frac{\Delta M_{j,t}^{OC}}{Y_{j,88} + M_{j,88} - E_{j,88}} \tag{10}$$

where $\Delta M_{j,t}^{OC}$ denotes the change in Chinese imports in industry $j$ over period $t$ for eight high-income countries other than the United States.[19] Autor et al. (2014) provide evidence on the robustness of this instrument in studying the impacts of the trade shock from China.[20] They argue that other high-income economies experience similar growth in Chinese imports that is driven by supply shocks originating in China, but are not exposed to the same demand shocks in the US. The instrumental variables, therefore, allow us to isolate the effect of exogenous supply-driven component of rising Chinese imports on ESG performance of U.S. firms.

---
[18] Occasionally, due to merger and acquisition, a firm has ESG scores earlier but not later, and, hence, is also excluded from the final sample.
[19] These countries are Australia, Denmark, Finland, Germany, Japan, New Zealand, Spain, and Switzerland.

Table 11 shows the impact of rising import penetration of China in the United States on the local firm's ESG score in the OLS model (column (1)-(2)), reduced-form (column (3)-(4)) and 2SLS model (column (5)-(6)). The result in column (2) shows that a one standard deviation increase in import penetration is associated with a 0.18-point increase in firm's ESG score, approximately 8% of a standard deviation of ESG scores.[21] We follow Autor et al. (2020) and assign zeros to import penetration of (non-manufacturing) industries that have no records of import from China. Column (2) adds a dummy of manufacturing sector (SIC 2000-3999), and shows that the result is not driven by industries with zero IPs.

Column (6) of Table 11 reports the result from 2SLS regression, and indicates a more prominent effect of import competition on ESG engagement. The magnitude of the coefficient more than doubles compared to the OLS regression, suggesting that IV estimation helps correct the measurement error in import penetration from China to the US. The result shows that one standard deviation increase in the trade exposure variable raises the firm's ESG score by 0.41 point, equivalent to 14% of one standard deviation of ESG scores. The estimated impact is very close to our estimate from the primary identification using PNTR. Therefore, using the alternative strategy by Autor et al. (2014), we find similar encouraging effects of the trade shock from China on US local firm's ESG engagement.

# VI. Conclusion

ESG performance starts to receive more attention from both corporates and investors. This paper exploits US congress granting Permanent Normal Trade Relations to China, and finds a causal link between greater import competition from China and rising ESG engagement by US firms. The better ESG performance mainly stems from firms "doing more positives" and from more involvement on environmental initiatives. The results are robust to the inclusion of firm, year, industry-year fixed effects and multiple firm-level covariates.

---

[20] Also see Hombert and Matray (2018) for importance to instrument the import penetration from China.
[21] Import penetration increased by a mean of 2.45 (standard deviation 9.49) for the firms in our sample.

We analyze two competing hypotheses. Our results indicate that the change in production process or offshoring by US companies is unlikely driving the improvement in ESG performance. Instead, we find both indirect and direct evidence supporting that US firms strategically become more involved in ESG initiatives in order to differentiate themselves from Chinese exporters.

The efficacy of such a strategy, nevertheless, is not clear. More generally, little is known about the impact, especially the long-term impact, of the engagement on ESG initiatives on a firm's performance. Figure 1 shows that our sample covers the nascent period of "ESG"/ "CSR" notion. Several recent studies (Fornell et al. (2006); Hong and Kacperczyk (2009); Edmans (2011); Bolton and Kacperczyk (2021)) find no positive relationship between ESG and stock returns, while others (Derwall et al. (2005); Kempf and Osthoff (2007); Statman and Glushkov (2009); Khan, Serafeim, and Yoon (2016)) present supporting evidence. We leave this question to future research.

# Reference

Acemoglu, Daron, David Autor, David Dorn, Gordon H Hanson, and Brendan Price, 2016, Import competition and the great US employment sag of the 2000s, *Journal of Labor Economics* 34, S141–S198.

Albuquerque, Ana M, Gus De Franco, and Rodrigo S Verdi, 2013, Peer choice in CEO compensation, *Journal of Financial Economics* 108, 160–181.

Auger, Pat, Paul Burke, Timothy M. Devinney, and Jordan J. Louviere, 2003, What Will Consumers Pay for Social Product Features?, *Journal of Business Ethics 2003 42:3* 42, 281–304.

Autor, David, David Dorn, and Gordon Hanson, 2019, When Work Disappears: Manufacturing Decline and the Falling Marriage Market Value of Young Men, *American Economic Review: Insights* 1, 161–78.

Autor, David, David Dorn, and Gordon Hanson, 2020, Importing Political Polarization? The Electoral Consequences of Rising Trade Exposure, *American Economic Review* 110, 3139–83.

Autor, David, David Dorn, and Gordon H. Hanson, 2016, The China Shock: Learning from Labor-Market Adjustment to Large Changes in Trade, *Annual Review of Economics* 8, 205–240.

Autor, David, David Dorn, Gordon H Hanson, Gary Pisano, and Pian Shu, 2020, Foreign competition and domestic innovation: Evidence from US patents, *American Economic Review:Insights* 2, 357–374.

Autor, David, David Dorn, Gordon H Hanson, and Jae Song, 2014, Trade adjustment: Worker-level evidence, *The Quarterly Journal of Economics* 129, 1799–1860.

Bagnoli, Mark, and Susan G. Watts, 2003, Selling to Socially Responsible Consumers: Competition and The Private Provision of Public Goods, *Journal of Economics & Management Strategy* 12, 419–445.

Ben-David, Itzhak, Yeejin Jang, Stefanie Kleimeier, and Michael Viehs, 2021, Exporting pollution: where do multinational firms emit $CO_2$?, *Economic Policy* 36, 377–437.

Bernard, Andrew B., J. Bradford Jensen, and Peter K. Schott, 2006, Survival of the best fit: Exposure to low-wage countries and the (uneven) growth of U.S. manufacturing plants, *Journal of International Economics* 68, 219–237.

Bizjak, John, Michael Lemmon, and Thanh Nguyen, 2011, Are all CEOs above average? An empirical analysis of compensation peer groups and pay design, *Journal of financial economics* 100, 538–555.

Bloom, Nicholas, Mirko Draca, and John Van Reenen, 2016, Trade Induced Technical Change? The Impact of Chinese Imports on Innovation, IT and Productivity, *The Review of Economic Studies* 83, 87–117.

Bolton, Patrick, and Marcin Kacperczyk, 2021, Do investors care about carbon risk?, *Journal of Financial Economics* 142, 517–549.

Borghesi, Richard, Joel F. Houston, and Andy Naranjo, 2014, Corporate socially responsible investments: CEO altruism, reputation, and shareholder interests, *Journal of Corporate*

*Finance* 26, 164–181.

Cai, Ye, Carrie H. Pan, and Meir Statman, 2016, Why do countries matter so much in corporate social performance?, *Journal of Corporate Finance* 41, 591–609.

Cao, Jie, Hao Liang, and Xintong Zhan, 2019, Peer Effects of Corporate Social Responsibility, *Management Science* 65, 5487–5503.

Chen, Tao, Hui Dong, and Chen Lin, 2020, Institutional shareholders and corporate social responsibility, *Journal of Financial Economics* 135, 483–504.

Creyer, Elizabeth H., and William T. Ross, 1997, The influence of firm behavior on purchase intention: Do consumers really care about business ethics?, *Journal of Consumer Marketing* 14, 421–430.

Cronqvist, Henrik, and Frank Yu, 2017, Shaped by their daughters: Executives, female socialization, and corporate social responsibility, *Journal of Financial Economics* 126, 543–562.

Da, Zhi, Joseph Engelberg, and Pengjie Gao, 2011, In Search of Attention, *The Journal of Finance* 66, 1461–1499.

David, Autor, David Dorn, and Gordon H Hanson, 2013, The China syndrome: Local labor market effects of import competition in the United States, *American Economic Review* 103, 2121–2168.

De Pelsmacker, Patrick, Liesbeth Driesen, and Glenn Rayp, 2005, Do Consumers Care about Ethics? Willingness to Pay for Fair-Trade Coffee, *Journal of Consumer Affairs* 39, 363–385.

Derwall, Jeroen, Nadja Guenster, Rob Bauer, and Kees Koedijk, 2005, The eco-efficiency premium puzzle, *Financial Analysts Journal* 61.

Dessaint, Olivier, Thierry Foucault, Laurent Frésard, and Adrien Matray, 2019, Noisy stock prices and corporate investment, *The Review of Financial Studies* 32, 2625–2672.

Di Giuli, Alberta, and Leonard Kostovetsky, 2014, Are red or blue companies more likely to go green? Politics and corporate social responsibility, *Journal of Financial Economics* 111, 158–180.

Dimson, Elroy, Oǧuzhan Karakaş, and Xi Li, 2015, Active Ownership, *The Review of Financial Studies* 28, 3225–3268.

Donangelo, Andres, François Gourio, Matthias Kehrig, and Miguel Palacios, 2019, The cross-section of labor leverage and equity returns, *Journal of Financial Economics* 132, 497–518.

Dyck, Alexander, Karl V. Lins, Lukas Roth, and Hannes F. Wagner, 2019, Do institutional investors drive corporate social responsibility? International evidence, *Journal of Financial Economics* 131, 693–714.

Edmans, Alex, 2011, Does the stock market fully value intangibles? Employee satisfaction and equity prices, *Journal of Financial Economics* 101, 621–640.

Elhanan, Helpman, and Paul Krugman, 1989, *Trade Policy and Market Structure* (MIT Press).

Esposito, Louis, and Frances Ferguson Esposito, 1971, Foreign Competition and Domestic

Industry Profitability, *The Review of Economics and Statistics* 53.

Faulkender, Michael, and Jun Yang, 2010, Inside the black box: The role and composition of compensation peer groups, *Journal of Financial Economics* 96, 257–270.

Fee, C. Edward, Charles J. Hadlock, and Joshua R. Pierce, 2009, Investment, Financing Constraints, and Internal Capital Markets: Evidence from the Advertising Expenditures of Multinational Firms, *The Review of Financial Studies* 22, 2361–2392.

Feenstra, Robert C., and Gordon H. Hanson, 1999, The impact of outsourcing and high-technology capital on wages: Estimates for the United States, 1979-1990, *Quarterly Journal of Economics* 114.

Flammer, Caroline, 2015, Does product market competition foster corporate social responsibility? Evidence from trade liberalization, *Strategic Management Journal* 36, 1469–1485.

Fornell, Claes, Sunil Mithas, Forrest V. Morgeson, and M. S. Krishnan, 2006, Customer satisfaction and stock prices: High returns, low risk, *Journal of Marketing* 70.

Foucault, Thierry, and Laurent Fresard, 2014, Learning from peers' stock prices and corporate investment, *Journal of Financial Economics* 111, 554–577.

Giannetti, Mariassunta, Mike Burkart, and Tore Ellingsen, 2011, What You Sell Is What You Lend? Explaining Trade Credit Contracts, *The Review of Financial Studies* 24, 1261–1298.

Gillan, Stuart L., Andrew Koch, and Laura T. Starks, 2021, Firms and social responsibility: A review of ESG and CSR research in corporate finance, *Journal of Corporate Finance* 66, 101889.

Gregory, Alan, Rajesh Tharyan, and Julie Whittaker, 2014, Corporate Social Responsibility and Firm Value: Disaggregating the Effects on Cash Flow, Risk and Growth, *Journal of Business Ethics* 124.

Grieser, William, Charles Hadlock, James LeSage, and Morad Zekhnini, 2021, Network effects in corporate financial policies, *Journal of Financial Economics*.

Gros, Daniel, 1987, A note on the optimal tariff, retaliation and the welfare loss from tariff wars in a framework with intra-industry trade, *Journal of International Economics* 23, 357–367.

Hasanbeigi, Ali, William Morrow, and Arman Shehabi, 2021, Embodied carbon in the US manufacturing and trade.

Hasanbeigi, Ali, Lynn Price, Zhang Chunxia, Nathaniel Aden, Li Xiuping, and Shangguan Fangqin, 2014, Comparison of iron and steel production energy use and energy intensity in China and the U.S., *Journal of Cleaner Production* 65.

Hegde, Shantaram P., and Dev R. Mishra, 2019, Married CEOs and corporate social responsibility, *Journal of Corporate Finance* 58, 226–246.

Hoberg, Gerard, and Gordon Phillips, 2010, Real and Financial Industry Booms and Busts, *The Journal of Finance* 65, 45–86.

Hoberg, Gerard, and Gordon Phillips, 2016, Text-based network industries and endogenous product differentiation, *Journal of Political Economy* 124, 1423–1465.

Hombert, Johan, and Adrien Matray, 2018, Can Innovation Help U.S. Manufacturing Firms Escape Import Competition from China?, *The Journal of Finance* 73, 2003–2039.

Hong, Harrison G, Jeffrey D Kubik, Inessa Liskovich, and José Scheinkman, 2019, Crime, punishment and the value of corporate social responsibility, *Available at SSRN 2492202*.

Hong, Harrison, and Marcin Kacperczyk, 2009, The price of sin: The effects of social norms on markets, *Journal of Financial Economics* 93, 15–36.

Hong, Harrison, Jeffrey D Kubik, and Jose A Scheinkman, 2012, Financial constraints on corporate goodness, National Bureau of Economic Research.

Hull, Clyde Eiríkur, and Sandra Rothenberg, 2008, Firm performance: the interactions of corporate social performance with innovation and industry differentiation, *Strategic Management Journal* 29, 781–789.

Iliev, Peter, and Lukas Roth, 2021, Do Directors Drive Corporate Sustainability?, *SSRN Electronic Journal*.

Jha, Anand, and James Cox, 2015, Corporate social responsibility and social capital, *Journal of Banking & Finance* 60, 252–270.

Kaustia, Markku, and Ville Rantala, 2015, Social learning and corporate peer effects, *Journal of Financial Economics* 117, 653–669.

Kelchtermans, Stijn, Daniel Neicu, and Peter Teirlinck, 2020, The role of peer effects in firms' usage of R&D tax exemptions, *Journal of Business Research* 108.

Kempf, Alexander, and Peer Osthoff, 2007, The effect of socially responsible investing on portfolio performance, *European Financial Management* 13.

Khan, Mozaffar, George Serafeim, and Aaron Yoon, 2016, Corporate sustainability: First evidence on materiality, *Accounting Review*.

Leary, Mark T, and Michael R Roberts, 2014, Do peer firms affect corporate financial policy?, *The Journal of Finance* 69, 139–178.

Li, Xuelin, Andrew W. Lo, and Richard T. Thakor, 2021, Paying Off the Competition: Market Power and Innovation Incentives, *SSRN Electronic Journal*.

Liang, Hao, and Luc Renneboog, 2017, On the Foundations of Corporate Social Responsibility, *The Journal of Finance* 72, 853–910.

Masulis, Ronald W., Cong Wang, and Fei Xie, 2009, Agency Problems at Dual-Class Companies, *The Journal of Finance* 64, 1697–1727.

Mion, Giordano, and Linke Zhu, 2013, Import competition from and offshoring to China: A curse or blessing for firms?, *Journal of International Economics* 89.

Moran, Daniel, Ali Hasanbeigi, and Cecilia Springer, 2018, The Carbon Loophole in Climate Policy: Quantifying the Embodied Carbon in Traded Products, *Report sponsored by KGM & Associates, Global Efficiency Intelligence, and ClimateWorks Foundation*.

Navarro, Peter, 1988, Why Do Corporations Give to Charity?, *The Journal of Business* 61, 65–93.

Peng, Zhen, Yujun Lian, and Joseph A Forson, 2021, Peer effects in R&D investment policy: Evidence from China, *International Journal of Finance & Economics* 26, 4516–4533.

Pierce, Justin R, and Peter K Schott, 2016, The surprisingly swift decline of US manufacturing employment, *American Economic Review* 106, 1632–1662.

Pugel, Thomas A., 1980, Foreign Trade and US Market Performance, *The Journal of Industrial Economics* 29.

Rauch, James E, 1999, Networks versus markets in international trade, *Journal of international Economics* 48, 7–35.

Schott, Peter K., 2008, The relative sophistication of Chinese exports, *Economic Policy*.

Siegel, Donald S., and Donald F. Vitaliano, 2007, An Empirical Analysis of the Strategic Use of Corporate Social Responsibility, *Journal of Economics & Management Strategy* 16, 773–792.

Starks, Laura T., Parth Venkat, and Qifei Zhu, 2017, Corporate ESG Profiles and Investor Horizons, *SSRN Electronic Journal*.

Statman, Meir, and Denys Glushkov, 2009, The wages of social responsibility, *Financial Analysts Journal* 65.

**Figures**

**Figure 1: ESG Related Google Search Volume Indices**

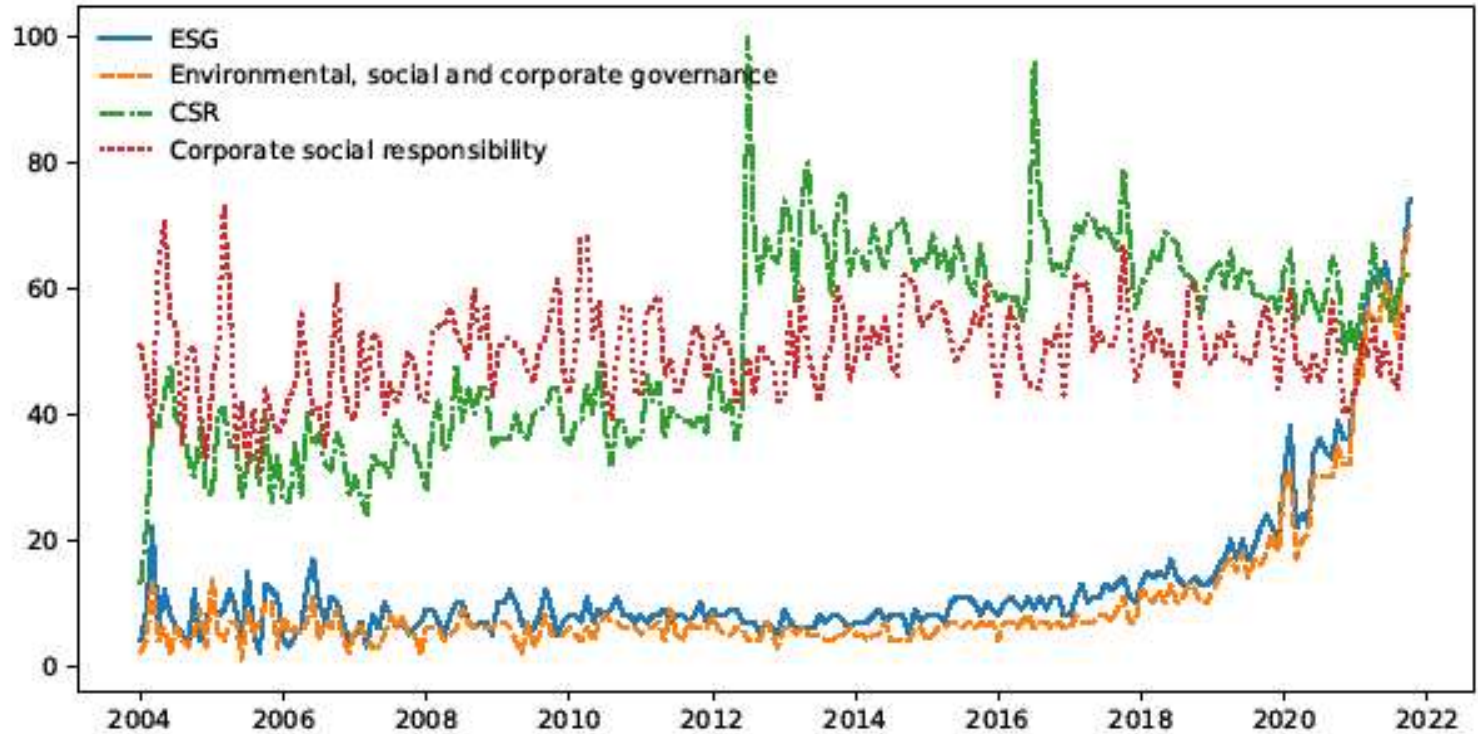

Figure 1 displays the search volume indices of ESG key words collected from Google Trend: ESG, Environmental, Social and corporate governance, CSR and Corporate Social Responsibility

**Figure 2: Job Satisfaction Survey Trends**

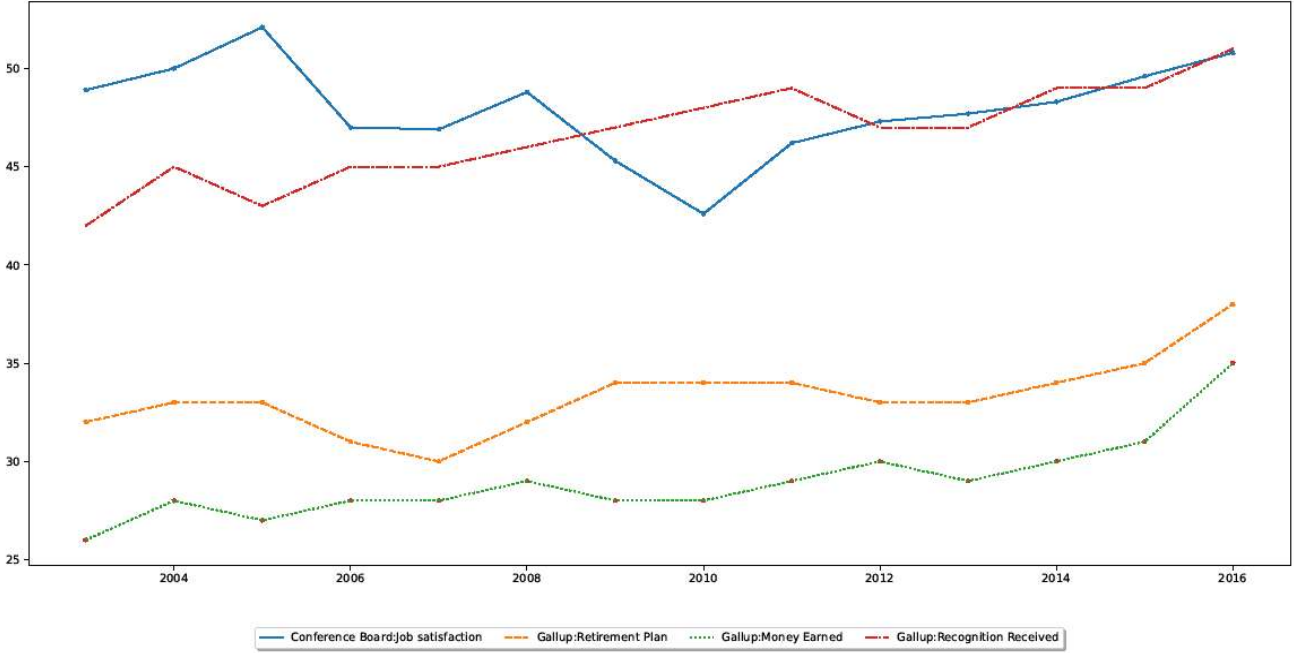

Figure 2 plots the results of job satisfactory survey by the Conference Board (solid), retirement-plan satisfaction survey by Gallup (dashed), money-earned satisfaction survey by Gallup (dotted) and recognition-received satisfaction survey by Gallup (dot-dashed) from 2003 to 2016.

**Figure 3: Estimated 95% Confidence Interval for DID Coefficients**

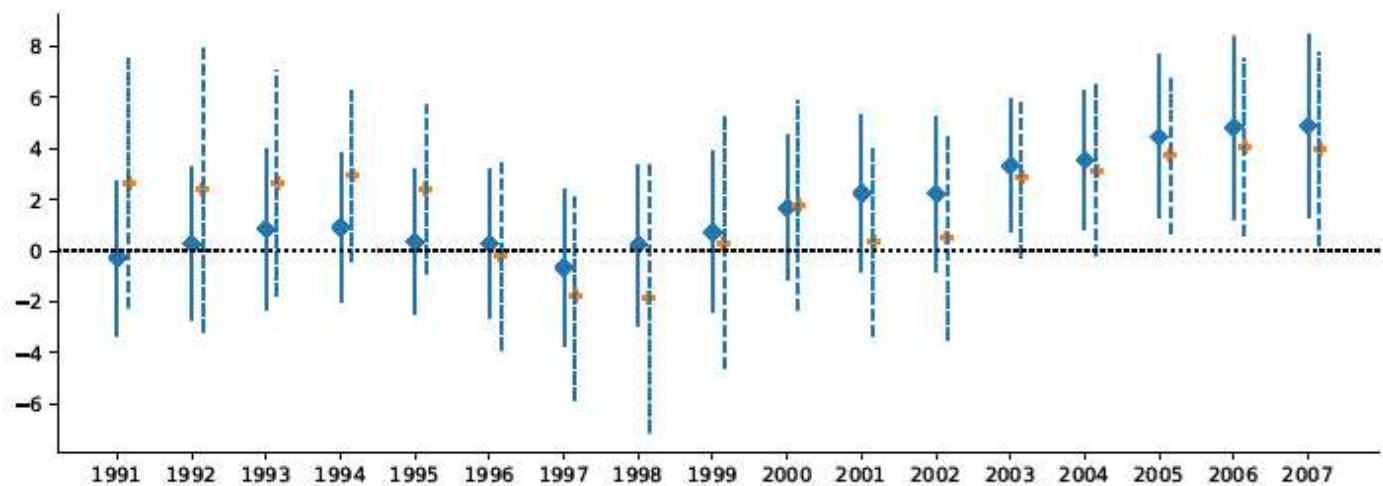

Figure 3 displays the 95 percent confidence interval for estimated DiD coefficients for interactions of year dummies with the NTR gap from equation (2). Solid lines represent the specification which includes only the DiD coefficients and the fixed effects. Dashed lines represent the specification which also controls for firm-level covariates. Firm-level covariates include (log) assets, return of assets (ROA), book to market ratio, cash, dividend and debt.

# Tables

## Table 1: Strengths and Concerns of Diversity in KLD

Table 1 shows indicators of strengths and concerns in KLD to evaluate a firm's performance on diversity. A firm obtains a score of 1 for a strength (concern) indicator if it performs well (poorly) in a particular criterion.

| Indicator | Description |
|---|---|
| Strength: Representation | at least one woman among the executive management team |
| Strength: Gender | with strong gender diversity on their board of directors |
| Concern: Discrimination & Workface Diversity | measures the severity of controversies related to a firm's workforce diversity, including its own employees as well as temporary employees, contractors, and franchisee employees. |
| Concern: Board Diversity - Gender | with no women on their board of directors |

# Table 2: Summary Statistics

Table 2 presents the summary statistics of ESG scores, individual constituent scores, and other firm-level covariates used in the baseline DiD regressions. Log(Asset) is the natural log of a firm's total assets (in $mil). ROA is return to assets. Book to Market is the ratio of book value of equity to market capitalization. Cash is the ratio of cash and short-term investments to total assets. Dividend is the ratio of dividend to total assets. Debt is total debt (short term and long term) scaled by total assets.

|  | Obs | Mean | Std Dev | P25 | P50 | P75 |
|---|---|---|---|---|---|---|
| ESG Score | 6,736 | -0.096 | 2.317 | -1 | 0 | 1 |
| Strength | 6,736 | 1.723 | 2.224 | 0 | 1 | 2 |
| Concern | 6,736 | 1.819 | 1.936 | 1 | 1 | 2 |
| E Score | 6,736 | -0.160 | 0.878 | 0 | 0 | 0 |
| S Score | 6,736 | 0.299 | 1.935 | -1 | 0 | 1 |
| G Score | 6,736 | -0.234 | 0.670 | -1 | 0 | 0 |
| NTR Gap | 6,731 | 0.282 | 0.149 | 0.193 | 0.325 | 0.377 |
| Log(Asset) | 6,576 | 7.140 | 1.733 | 5.906 | 7.098 | 8.315 |
| ROA | 6,575 | 0.013 | 0.194 | 0.004 | 0.048 | 0.090 |
| Book to Market | 6,567 | 0.411 | 0.535 | 0.194 | 0.328 | 0.530 |
| Cash | 6,576 | 0.204 | 0.234 | 0.027 | 0.105 | 0.303 |
| Dividend | 6,563 | 0.014 | 0.044 | 0 | 0.004 | 0.019 |
| Debt | 6,555 | 0.206 | 0.204 | 0.044 | 0.186 | 0.311 |

**Table 3: The Effects of Chinese Import Competition on Firms' ESG Performance**

Table 3 reports results of DiD regressions of ESG scores on the interaction of NTR gaps with an indicator for the Post_PNTR period. Log(Asset) is the natural log of a firm's total assets (in $mil). ROA is return to assets. Book to Market is the ratio of book value of equity to market capitalization. Cash is the ratio of cash and short-term investments to total assets. Dividend is the ratio of dividend to total assets. Debt is total debt (short term and long term) scaled by total assets. Standard errors are clustered at 4-digit SIC level. ***, **, * denote significance at the 1%, 5%, and 10% levels, respectively.

|  | ESG Scores | | | |
|---|---|---|---|---|
|  | (1) | (2) | (3) | (4) |
| Post_PNTR×NTR Gap | 1.316** | 1.790*** | 3.104*** | 1.909*** |
|  | (0.554) | (0.677) | (1.053) | (0.695) |
| Log(Asset) | -0.203 | -0.030 | -0.119 | -0.125 |
|  | (0.181) | (0.104) | (0.169) | (0.193) |
| ROA | 0.422** | 0.813*** | 0.442*** | 0.462** |
|  | (0.177) | (0.233) | (0.171) | (0.191) |
| Book to Market | -0.022 | -0.236*** | -0.058 | -0.119 |
|  | (0.087) | (0.090) | (0.091) | (0.102) |
| Cash | 0.112 | 0.521** | 0.212 | 0.052 |
|  | (0.283) | (0.245) | (0.284) | (0.334) |
| Dividend | -0.436 | 1.880** | -0.523 | -0.497 |
|  | (0.692) | (0.894) | (0.683) | (0.772) |
| Debt | 0.404 | -0.501* | 0.283 | -0.199 |
|  | (0.295) | (0.302) | (0.274) | (0.311) |
| Firm Fixed Effects | Yes | No | Yes | Yes |
| Year Fixed Effects | No | Yes | Yes | Yes |
| Industry×Year Fixed Effects | No | No | No | Yes |
| Observations | 6,528 | 6,528 | 6,528 | 6,528 |
| $R^2$ | 0.709 | 0.060 | 0.714 | 0.778 |

# Table 4: The Effects of Chinese Import Competition on Firms' ESG Performance: Constituents

Table 4 reports results of DiD regressions of ESG constituent scores on the interaction of the NTR gaps with an indicator for the Post_PNTR period. Log(Asset) is the natural log of a firm's total assets (in $mil). ROA is return to assets. Cash is the ratio of cash and short-term investments to total assets. Dividend is the ratio of dividend to total assets. Debt is total debt (short term and long term) scaled by total assets. All specifications control for firm, year, and industry-year fixed effects. Standard errors are clustered at 4-digit SIC level. ***, **, * denote significance at the 1%, 5%, and 10% levels, respectively.

|  | Strengths (1) | Concerns (2) | E-scores (3) | S-scores (4) | G-scores (5) |
|---|---|---|---|---|---|
| Post_PNTR×NTR Gap | 1.894** | -0.015 | 1.023*** | 0.742 | 0.143 |
|  | (0.898) | (0.878) | (0.301) | (0.758) | (0.299) |
| Log(Asset) | 0.066 | 0.190** | -0.104** | 0.187 | -0.208*** |
|  | (0.149) | (0.087) | (0.047) | (0.155) | (0.038) |
| ROA | 0.133 | -0.329** | -0.016 | 0.417** | 0.061 |
|  | (0.132) | (0.143) | (0.046) | (0.178) | (0.075) |
| Book to Market | -0.085 | 0.033 | 0.001 | -0.103 | -0.017 |
|  | (0.063) | (0.054) | (0.036) | (0.070) | (0.023) |
| Cash | 0.090 | 0.038 | 0.136 | -0.042 | -0.042 |
|  | (0.375) | (0.197) | (0.099) | (0.326) | (0.124) |
| Dividend | -0.798 | -0.301 | -0.164 | 0.141 | -0.475** |
|  | (0.743) | (0.281) | (0.199) | (0.625) | (0.225) |
| Debt | -0.040 | 0.159 | -0.057 | -0.149 | 0.007 |
|  | (0.214) | (0.229) | (0.098) | (0.311) | (0.086) |
| Firm Fixed Effects | Yes | Yes | Yes | Yes | Yes |
| Year Fixed Effects | Yes | Yes | Yes | Yes | Yes |
| Industry×Year Fixed Effects | Yes | Yes | Yes | Yes | Yes |
| Observations | 6,528 | 6,528 | 6,528 | 6,528 | 6,528 |
| $R^2$ | 0.849 | 0.849 | 0.800 | 0.794 | 0.643 |

# Table 5: The Effects of Chinese Import Competition on Firms' ESG Performance Change in Production Process

Table 5 reports results of DiD regressions of ESG scores on the interaction of the NTR gaps with an indicator for the Post_PNTR period. Log(Asset) is the natural log of a firm's total assets (in $mil). ROA is return to assets. Cash is the ratio of cash and short-term investments to total assets. Dividend is the ratio of dividend to total assets. Debt is total debt (short term and long term) scaled by total assets. Staff expense is the ratio of staff expense to total sales. Capital intensity is the ratio of capital expenditure to total number of employees. R&D expense is R&D expenditure in Compustat scaled by total sales. Advertising is the ratio of advertising expenditure to total sales. Missing values of staff expense, R&D expense and advertising expenditure are replaced with zeros. All specifications control for firm, year, and industry-year fixed effects. Standard errors are clustered at 4-digit SIC level. ***, **, * denote significance at the 1%, 5%, and 10% levels, respectively.

|  | ESG Scores | | | | |
| --- | --- | --- | --- | --- | --- |
|  | (1) | (2) | (3) | (4) | (5) |
| Post_PNTR×NTR Gap | 1.871*** | 1.868*** | 1.984*** | 1.896*** | 1.825*** |
|  | (0.715) | (0.682) | (0.696) | (0.679) | (0.706) |
| Staff Expense | -3.359 |  |  |  | -3.336 |
|  | (2.254) |  |  |  | (2.199) |
| Post_PNTR×Staff Expense | 1.227 |  |  |  | 1.210 |
|  | (1.533) |  |  |  | (1.491) |
| Capital Intensity |  | 0.001 |  |  | 0.001 |
|  |  | (0.001) |  |  | (0.001) |
| Post_PNTR×Capital Intensity |  | -0.001 |  |  | -0.001 |
|  |  | (0.001) |  |  | (0.001) |
| R&D Expense |  |  | -2.400 |  | -2.545 |
|  |  |  | (2.886) |  | (2.972) |
| Post_PNTR×R&D |  |  | 2.400 |  | 2.545 |
|  |  |  | (2.886) |  | (2.972) |
| Advertising |  |  |  | -1.468 | -0.371 |
|  |  |  |  | (4.480) | (4.335) |
| Post_PNTR×Advertising |  |  |  | 1.596 | 3.758 |
|  |  |  |  | (4.732) | (5.906) |
| Log(Asset) | -0.112 | -0.099 | -0.141 | -0.124 | -0.103 |
|  | (0.192) | (0.196) | (0.198) | (0.192) | (0.203) |
| ROA | 0.470** | 0.469* | 0.476** | 0.461** | 0.509** |
|  | (0.190) | (0.243) | (0.196) | (0.189) | (0.255) |
| Book to Market | -0.123 | -0.052 | -0.116 | -0.118 | -0.056 |
|  | (0.097) | (0.071) | (0.101) | (0.101) | (0.068) |
| Cash | 0.070 | 0.075 | 0.048 | 0.057 | 0.093 |
| | (0.333) | (0.338) | (0.326) | (0.331) | (0.335) |
| Dividend | -0.513 | -0.577 | -0.484 | -0.478 | -0.512 |
| | (0.760) | (0.773) | (0.768) | (0.773) | (0.757) |
| Debt | -0.192 | -0.134 | -0.195 | -0.198 | -0.121 |
| | (0.311) | (0.312) | (0.310) | (0.310) | (0.318) |
| Firm Fixed Effects | Yes | Yes | Yes | Yes | Yes |
| Year Fixed Effects | Yes | Yes | Yes | Yes | Yes |
| Industry×Year Fixed Effects | Yes | Yes | Yes | Yes | Yes |
| Observations | 6,528 | 6,450 | 6,528 | 6,528 | 6,450 |
| $R^2$ | 0.779 | 0.780 | 0.779 | 0.778 | 0.782 |

# Table 6: Effect of Chinese Import Competition on Firms' ESG Performance Change in Production Process

Table 6 reports results of DiD regressions of ESG scores on the interaction of the NTR gaps with an indicator for the *Post_PNTR* period. Offshoring is the sum product of the input weights and import penetration of other industries. Other variables are defined in the same way as Table 5 except staff expense, capital intensity, R&D expense, and advertising expense are averages of firms in the same 4-digit SIC industries. All specifications control for firm, year, and industry-year fixed effects. Standard errors are clustered at 4-digit SIC level. ***, **, * denote significance at the 1%, 5%, and 10% levels, respectively.

|  | ESG Scores | | | | | |
|---|---|---|---|---|---|---|
|  | (1) | (2) | (3) | (4) | (5) | (6) |
| Post_PNTR×NTR Gap | 2.200* | 2.439*** | 1.766** | 1.999*** | 2.005*** | 3.718** |
|  | (1.229) | (0.706) | (0.731) | (0.712) | (0.727) | (1.524) |
| Staff Expense | -0.026*** |  |  |  |  | -0.027*** |
|  | (0.003) |  |  |  |  | (0.002) |
| Post_PNTR×Staff Expense | 0.033*** |  |  |  |  | 0.034*** |
|  | (0.003) |  |  |  |  | (0.006) |
| Capital Intensity |  | -0.027 |  |  |  | -0.047 |
|  |  | (0.017) |  |  |  | (0.044) |
| Post_PNTR×Capital Intensity |  | 0.056*** |  |  |  | 0.099*** |
|  |  | (0.021) |  |  |  | (0.038) |
| RD Expense |  |  | -0.007 |  |  | -0.038 |
|  |  |  | (0.019) |  |  | (0.027) |
| Post_PNTR×R & D |  |  | 0.007 |  |  | 0.038 |
|  |  |  | (0.019) |  |  | (0.026) |
| Advertising |  |  |  | -0.031 |  | -0.741** |
|  |  |  |  | (0.058) |  | (0.334) |
| Post_PNTR×Advertising |  |  |  | 0.207 |  | 0.390 |
|  |  |  |  | (0.165) |  | (0.417) |
| Offshoring |  |  |  |  | -0.009 | 0.088 |
|  |  |  |  |  | (0.024) | (0.055) |
| Post_PNTR×Offshoring |  |  |  |  | 0.014 | -0.040 |
|  |  |  |  |  | (0.021) | (0.047) |
| Log(Asset) | -0.240 | -0.130 | -0.142 | -0.126 | -0.073 | -0.182 |
|  | (0.157) | (0.194) | (0.198) | (0.193) | (0.227) | (0.189) |
| ROA | 0.546** | 0.443** | 0.483** | 0.456** | 0.405** | 0.433** |
|  | (0.212) | (0.186) | (0.194) | (0.186) | (0.184) | (0.173) |
| Book to Market | -0.099 | -0.117 | -0.128 | -0.120 | -0.136 | -0.147 |

|  | (0.109) | (0.102) | (0.103) | (0.102) | (0.107) | (0.111) |
|---|---|---|---|---|---|---|
| Cash | 0.253 | 0.028 | 0.031 | 0.014 | -0.004 | 0.226 |
|  | (0.297) | (0.331) | (0.334) | (0.325) | (0.363) | (0.323) |
| Dividend | -0.308 | -0.550 | -0.516 | -0.560 | -0.542 | -0.246 |
|  | (0.938) | (0.784) | (0.784) | (0.795) | (0.824) | (0.928) |
| Debt | -0.192 | -0.191 | -0.192 | -0.120 | -0.161 | -0.028 |
|  | (0.348) | (0.313) | (0.318) | (0.296) | (0.339) | (0.357) |
| Firm Fixed Effects | Yes | Yes | Yes | Yes | Yes | Yes |
| Year Fixed Effects | Yes | Yes | Yes | Yes | Yes | Yes |
| Industry×Year Fixed Effects | Yes | Yes | Yes | Yes | Yes | Yes |
| Observations | 4,940 | 6,528 | 6,270 | 6,195 | 5,651 | 4,017 |
| $R^2$ | 0.775 | 0.780 | 0.778 | 0.779 | 0.772 | 0.770 |

# Table 7: Effect of Chinese Import Competition on Firms' ESG Performance Politics & Headquarter Locations

Table 7 reports results of triple-difference regression of ESG scores on the interaction of the *NTR gaps*, an indicator (*Post_PNTR*) for the post PNTR period and an a variable measuring the leaning of a firm's headquarter state to Republican. *% Rep Delegates* is the average percentage of Republican delegates of the state between 2000 and 2006. *% Rep Votes* is the average percentage of state electorates voting for the Republican presidential candidates during the presidential elections in 2000 and 2004. *Red_congress* is a dummy variable equal to 1 if the average percentage of Republican delegates in the Congress is greater than the average percentage of Democratic delegates. *Red_president* is a dummy variable equal to 1 if the average percentage of state electorates voting for the Republican candidate is greater than the average percentage voting for the Democratic candidate. Other (not shown) controls include: log(Asset) is the natural log of a firm's total assets (in $mil); ROA is return to assets; cash is the ratio of cash and short-term investments to total assets; dividend is the ratio of dividend to total assets; debt is total debt (short term and long term) scaled by total assets. All specifications control for firm, year, and industry-year fixed effects. Standard errors are clustered at 4-digit SIC level. ***, **, * denote significance at the 1%, 5%, and 10% levels, respectively.

|  | ESG Scores | | | |
| --- | --- | --- | --- | --- |
|  | (1) | (2) | (3) | (4) |
| Post_PNTR×NTR Gap× % Rep Delegates | 0.091 | | | |
|  | (0.063) | | | |
| Post_PNTR×NTR Gap×Red_congress | | 5.385** | | |
|  | | (2.700) | | |
| Post_PNTR×NTR Gap× % Rep Votes | | | 0.323*** | |
|  | | | (0.091) | |
| Post_PNTR×NTR Gap×Red_president | | | | 5.945*** |
|  | | | | (1.478) |
| Post_PNTR×NTR Gap | -1.405 | 0.901 | -11.798*** | 1.017 |
|  | (2.389) | (0.816) | (3.867) | (0.750) |
| Full Controls | Yes | Yes | Yes | Yes |
| Firm Fixed Effects | Yes | Yes | Yes | Yes |
| Year Fixed Effects | Yes | Yes | Yes | Yes |
| Industry×Year Fixed Effects | Yes | Yes | Yes | Yes |
| Observations | 6,276 | 6,528 | 6,276 | 6,528 |
| $R^2$ | 0.784 | 0.780 | 0.785 | 0.781 |

## Table 8: China Trade Shock, Market Power and ESG Engagement

Table 8 reports results of triple-difference regression of ESG scores on the interaction of the NTR gaps, an indicator for the post PNTR period, and an indicator for industries with high Herfindahl-Hirschman Index (HHI). Other (not shown) controls include: log(Asset) is the natural log of a firm's total assets (in $mil); ROA is return to assets; cash is the ratio of cash and short-term investments to total assets; dividend is the ratio of dividend to total assets; debt is total debt (short term and long term) scaled by total assets. All specifications control for firm, year, and industry-year fixed effects. Standard errors are clustered at 4-digit SIC level. ***, **, * denote significance at the 1%, 5%, and 10% levels, respectively.

|  | ESG Scores | | | |
|---|---|---|---|---|
|  | (1) | (2) | (3) | (4) |
| Post_PNTR×NTR Gap×High_HHI (Hoberg &Phillips (2010)) | -0.609 | -8.565*** |  |  |
|  | (2.982) | (2.768) |  |  |
| Post_PNTR×NTR Gap×High_HHI (Hoberg &Phillips (2016)) |  |  | -3.987** | -2.917 |
|  |  |  | (1.990) | (2.177) |
| Post_PNTR×NTR Gap | 3.132*** | 2.192*** | 5.328*** | 4.222* |
|  | (0.783) | (0.704) | (1.735) | (2.167) |
| Full Controls | Yes | Yes | Yes | Yes |
| Firm Fixed Effects | Yes | Yes | Yes | Yes |
| Year Fixed Effects | Yes | Yes | Yes | Yes |
| Industry×Year Fixed Effects | No | Yes | No | Yes |
| Observations | 6,249 | 6,249 | 6,343 | 6,343 |
| $R^2$ | 0.711 | 0.773 | 0.710 | 0.773 |

## Table 9: Trade Shock, Differentiability and ESG Engagement

Table 9 reports the results of triple-difference regression of ESG scores on the interaction of the NTR gaps, an indicator for the post PNTR period, and an indicator for standardized industries. Other (not shown) controls include: log(Asset) is the natural log of a firm's total assets (in $mil); ROA is return to assets; cash is the ratio of cash and short-term investments to total assets; dividend is the ratio of dividend to total assets; debt is total debt (short term and long term) scaled by total assets. All specifications control for firm, year, and industry-year fixed effects. Standard errors are clustered at 4-digit SIC level. ***, **, * denote significance at the 1%, 5%, and 10% levels, respectively.

|  | ESG Scores | | | |
| --- | --- | --- | --- | --- |
|  | (1) | (2) | (3) | (4) |
| Post_PNTR×NTR Gap×Standarized (Hoberg &Phillips (2016)) | 3.897* | 1.695 | | |
|  | (2.112) | (1.085) | | |
| Post_PNTR×NTR Gap×Standarized (Rauch (1999)) | | | 0.806 | 5.312*** |
|  | | | (2.159) | (1.366) |
| Post_PNTR×NTR Gap | 1.181 | 0.840* | 2.154** | 1.803 |
|  | (1.539) | (0.500) | (0.899) | (1.265) |
| Full Controls | Yes | Yes | Yes | Yes |
| Firm Fixed Effects | Yes | Yes | Yes | Yes |
| Year Fixed Effects | Yes | Yes | Yes | Yes |
| Industry×Year Fixed Effects | No | Yes | No | Yes |
| Observations | 6,343 | 6,343 | 5,168 | 5,168 |
| $R^2$ | 0.712 | 0.773 | 0.718 | 0.781 |

# Table 10: Trade Shock from China & Firms' ESG Performance: Robustness Check

Table 10 reports results of DiD regressions of ESG scores on the interaction of NTR gaps with an indicator for the Post_PNTR period. Column (2) and (3) instrument NTR gaps with no-NTR tariff rates set by Smoot–Hawley Tariff Act and the *NTR gaps* observed in **1990,** respectively. Log(Asset) is the natural log of a firm's total assets (in $mil). ROA is return to assets. Book to Market is the ratio of book value of equity to market capitalization. Cash is the ratio of cash and short-term investments to total assets. Dividend is the ratio of dividend to total assets. Debt is total debt (short term and long term) scaled by total assets. Standard errors are clustered at 4-digit SIC level. ***, **, * denote significance at the 1%, 5%, and 10% levels, respectively.

|  | ESG Scores | | |
| --- | --- | --- | --- |
|  | (1) | (2) | (3) |
| Post_PNTR×NTR Gap | 1.909*** | 1.692** | 2.332** |
|  | (0.695) | (0.666) | (1.036) |
| Log(Asset) | -0.125 | -0.126 | -0.122 |
|  | (0.193) | (0.193) | (0.193) |
| ROA | 0.462** | 0.463** | 0.460** |
|  | (0.191) | (0.192) | (0.191) |
| Book to Market | -0.119 | -0.118 | -0.119 |
|  | (0.102) | (0.102) | (0.101) |
| Cash | 0.052 | 0.054 | 0.045 |
|  | (0.334) | (0.334) | (0.334) |
| Dividend | -0.497 | -0.498 | -0.496 |
|  | (0.772) | (0.772) | (0.771) |
| Debt | -0.199 | -0.199 | -0.198 |
|  | (0.311) | (0.311) | (0.311) |
| Firm Fixed Effects | Yes | Yes | Yes |
| Year Fixed Effects | Yes | Yes | Yes |
| Industry×Year Fixed Effects | Yes | Yes | Yes |
| Instrument Variable | - | Non-NTR tariff rates | NTR Gap (1990) |
| Observations | 6,528 | 6,528 | 6,518 |
| $R^2$ | 0.778 | 0.778 | 0.778 |

# Table 11: Chinese Import Competition & Firms' ESG Performance: Alternative Identification

Table 11 reports the results of OLS, reduced-form and 2SLS IV regressions of change in ESG scores on change in Chinese import penetration. The sample consist of two stacked long-difference sub-periods, 1991 to 1999 and 1999 to 2007. Log(Asset) is the natural log of a firm's total assets (in $mil). ROA is return to assets. Cash is the ratio of cash and short-term investments to total assets. Dividend is the ratio of dividend to total assets. Debt is total debt (short term and long term) scaled by total assets. Standard errors are clustered at 4-digit SIC level. ***, **, * denote significance at the 1%, 5%, and 10% levels, respectively.

|  | ESG Scores | | | | | |
|---|---|---|---|---|---|---|
|  | US imports from China (OLS) | | Third country imports from China (OLS reduced form) | | Third country imports from China (2SLS) | |
|  | (1) | (2) | (3) | (4) | (5) | (6) |
| Exposure variable | 0.035*** | 0.019** | 0.064*** | 0.044** | 0.060*** | 0.043** |
|  | (0.007) | (0.007) | (0.023) | (0.016) | (0.018) | (0.017) |
| Log Asset |  | 0.123 |  | 0.126 |  | 0.114 |
|  |  | (0. 230) |  | (0.232) |  | (0.228) |
| ROA |  | -0.971 |  | -0.970 |  | -1.047 |
|  |  | (1.901) |  | (1.835) |  | (1.915) |
| Book to Market |  | -0.074 |  | -0.108 |  | -0.094 |
|  |  | (0.272) |  | (0.266) |  | (0.271) |
| Cash |  | -0.811 |  | -1.159 |  | -0.971 |
|  |  | (1.189) |  | (1.315) |  | (1.240) |
| Dividend |  | 8.855 |  | 8.217 |  | 8.512 |
|  |  | (6.108) |  | (5.861) |  | (6.12) |
| Debt |  | -0.241 |  | -0.282 |  | -0.181 |
|  |  | (1.018) |  | (1.042) |  | (1.037) |
| Manufacture sector dummy |  | Yes |  | Yes |  | Yes |
| Observations | 565 | 553 | 565 | 553 | 565 | 553 |
| $R^2$ | 0.018 | 0.057 | 0.027 | 0.063 | 0.011 | 0.051 |

# Appendix:

## Table A.1: Categories and Indicators included in the ESG measures

| Category | Indicator | Strength/Concern |
|---|---|---|
| Environment | Environmental Opportunities – Opportunities in Clean Tech | Strength |
| Environment | Pollution & Waste – Toxic Emissions and Waste | Strength |
| Environment | Pollution & Waste – Packaging Materials & Waste | Strength |
| Environment | Climate Change - Carbon Emissions | Strength |
| Environment | Environmental Management Systems | Strength |
| Environment | Natural Capital - Water Stress | Strength |
| Environment | Natural Capital - Biodiversity & Land Use | Strength |
| Environment | Natural Capital - Raw Material Sourcing | Strength |
| Environment | Climate change - Financing Environmental Impact | Strength |
| Environment | Environmental Opportunities – Opportunities in Green Building | Strength |
| Environment | Environmental Opportunities – Opportunities in Renewable Energy | Strength |
| Environment | Pollution & Waste - Electronic Waste | Strength |
| Environment | Climate Change – Energy Efficiency | Strength |
| Environment | Climate Change – Product Carbon Footprint | Strength |
| Environment | Climate Change - Climate Change Vulnerability | Strength |
| Environment | Environment - Other Strengths | Strength |
| Environment | Toxic Emissions and Waste | Concern |
| Environment | Energy & Climate Change | Concern |
| Environment | Biodiversity & Land Use | Concern |
| Environment | Operational Waste (Non-Hazardous) | Concern |
| Environment | Supply Chain Management | Concern |
| Environment | Water Stress | Concern |
| Environment | Environment - Other Concerns | Concern |

| Category | Indicator | Strength/Concern |
|---|---|---|
| Social | Community Engagement | Strength |
| Social | Impact on Community | Concern |
| Social | Union Relations | Strength |
| Social | Cash Profit Sharing | Strength |
| Social | Employee Involvement | Strength |
| Social | Employee Health & Safety | Strength |
| Social | Supply Chain Labor Standards | Strength |
| Social | Human Capital Development | Strength |
| Social | Labor Management | Strength |
| Social | Controversial Sourcing | Strength |
| Social | Human Capital – Other Strengths | Strength |
| Social | Collective Bargaining & Unions | Concern |
| Social | Health & Safety | Concern |
| Social | Supply Chain Labor Standards | Concern |
| Social | Child Labor | Concern |
| Social | Labor Management Relations | Concern |
| Social | Labor Rights & Supply Chain – Other Concerns | Concern |
| Social | Representation | Strength |
| Social | Board Diversity - Gender | Strength |
| Social | Discrimination & Workforce Diversity | Concern |
| Social | Board Diversity - Gender | Concern |
| Social | Product Safety and Quality | Strength |
| Social | Social Opportunities – Access to Healthcare | Strength |
| Social | Social Opportunities - Access to Finance | Strength |
| Social | Social Opportunities - Access to Communications | Strength |
| Social | Social Opportunities - Opportunities in Nutrition and Health | Strength |
| Social | Product Safety - Chemical Safety | Strength |
| Social | Product Safety -Financial Product Safety | Strength |
| Social | Product Safety - Privacy & Data Security | Strength |
| Social | Product Safety - Responsible Investment | Strength |
| Social | Product Safety - Insuring Health and Demographic Risk | Strength |
| Social | Product Quality & Safety | Concern |
| Social | Marketing & Advertising | Concern |
| Social | Anticompetitive Practices | Concern |
| Social | Customer Relations | Concern |
| Social | Privacy & Data Security | Concern |
| Social | Other Concerns | Concern |

| Category | Indicator | Strength/Concern |
|---|---|---|
| Governance | Corruption & Political Instability | Strength |
| Governance | Financial System Instability | Strength |
| Governance | Governance Structures | Concern |
| Governance | Controversial Investments | Concern |
| Governance | Bribery & Fraud | Concern |
| Governance | Governance - Other Concerns | Concern |

## Table A.2: Variable Definitions and Descriptions

| Variable | Variable Description | Data Source |
| --- | --- | --- |
| ESG Scores | the difference between total strengths and total concerns | KLD |
| Total Strengths | total strengths | KLD |
| Total Concerns | total concerns | KLD |
| Environmental Scores | the difference between total environmental strenghts and environmental concerns | KLD |
| Social Scores | the difference between total social strengths and social concerns | KLD |
| Governance Scores | the difference between total governance strengths and governance concerns | KLD |
| Log(Asset) | natural log of a firm's total assets (AT) | Compustat |
| ROA | return to assets. Income Before Extraordinary Items (IB) over total assets (AT) | Compustat |
| Book to Market | Book to Market ratio. Book value of equity is CEQ item in Compustat. Market value of equity is the product of share price (PRC) and shares outstanding (SHROUT) in CRSP. | Compustat & CRSP |
| Cash | Cash and Short-Term Investments (CHE) over total assets (AT) | Compustat |
| Dividend | Dividend over total assets. Dividend is the sum of common dividend (DVC) and preferred dividend (DVP) | Compustat |
| Debt | Total debt outstanding over total assets. Debt is the sum of short-term debt (dlc) and long-term debt (dltt) | Compustat |
| Staff Expense (firm level) | Staff expense (XLR) over sales (SALE). Missing values of XLR are replaced with 0. | Compustat |
| Capital Intensity (firm level) | Capital expenditure (CAPX) over total number of employees (EMP) | Compustat |

**Variable Definitions and Descriptions (Cont'd)**

| Variable | Variable Description | Data Source |
|---|---|---|
| R&D Expense (firm level) | R&D expenditure (XRD) over sales (SALE). Missing values of XRD are replaced with 0. | Compustat |
| Advertising (firm level) | Advertising expenditure (XAD) over sales (SALE). Missing values of XAD are replaced with 0. | Compustat |
| Staff Expense (industry level) | average of firm-level staff expense for all firms in the same 4-digit SIC industries in the same year. Ignore any missing values. | Compustat |
| Capital Intensity (industry level) | average of firm-level capital intensity for all firms in the same 4-digit SIC industries in the same year. Ignore any missing values. | Compustat |
| R&D Expense (industry level) | average of firm-level R&D expenditure for all firms in the same 4-digit SIC industries in the same year. Ignore any missing values. | Compustat |
| Advertising (industry level) | average of firm-level advertising expenditure for all firms in the same 4-digit SIC industries in the same year. Ignore any missing values. | Compustat |
| % Rep Delegates | the average percentage of Republican delegates of the state between 2000 and 2006. | Ballotpedia |
| Red_congress | a dummy variable equal to 1 if the average percentage of Republican delegates in the Congress is greater than the average percentage of Democratic delegates | Ballotpedia |
| % Rep Votes | the average percentage of state electorates voting for the Republican presidential candidates during the presidential elections in 2000 and 2004. | Ballotpedia |
| Red_president | a dummy variable equal to 1 if the average percentage of state electorates voting for the Republican candidate is greater than the average percentage voting for the Democratic candidate | Ballotpedia |

**Variable Definitions and Descriptions (Cont'd)**

| Variable | Variable Description | Data Source |
|---|---|---|
| High_HHI | a dummy variable equal to 1 if the calculated industry HHI is above the median; based on Hoberg and Phillips (2010) and Hoberg and Phillips (2016) | Hoberg-Phillips Data Library |
| Standardized | a dummy variable equal to 1 if the 3-digit industry is classified as standardized based on product similarity measure of Hoberg and Phillips (2016) or Rauch (1999) | Hoberg-Phillips Data Library; James Rauch's website |
| Input-Output Matrix | a matrix of weights$_{k,j}$ of inputs in industry $k$ needed to produce one unit of final good in industry $j$ | BEA |
| Import Penetration | imports from China scaled by the initial absorption level at the start of the period in 1991 | Acemoglu et al. (2016) |

## Table A.3: Trade Shock, Differentiability and ESG Engagement: Liberal Classification of Standardized Industries

Table A.3 reports the results of triple-difference regression of ESG scores on the interaction of the NTR gaps, an indicator for the post PNTR period, and an indicator for standardized industries based on liberal classification in Rauch (1999). Other (not shown) controls include: log(Asset) is the natural log of a firm's total assets (in $mil); ROA is return to assets; cash is the ratio of cash and short-term investments to total assets; dividend is the ratio of dividend to total assets; debt is total debt (short term and long term) scaled by total assets. All specifications control for firm, year, and industry-year fixed effects. Standard errors are clustered at 4-digit SIC level. ***, **, * denote significance at the 1%, 5%, and 10% levels, respectively.

|  | ESG Scores | |
| --- | --- | --- |
|  | (1) | (2) |
| Post_PNTR×NTR Gap×Standarized (Rauch (1999)) | 1.348 | 5.360*** |
|  | (2.220) | (1.576) |
| Post_PNTR×NTR Gap | 2.193** | 1.817 |
|  | (0.906) | (1.247) |
| Full Controls | Yes | Yes |
| Firm Fixed Effects | Yes | Yes |
| Year Fixed Effects | Yes | Yes |
| Industry×Year Fixed Effects | No | Yes |
| Observations | 5,168 | 5,168 |
| $R^2$ | 0.717 | 0.781 |